\documentclass[sigconf]{acmart}

\usepackage{graphicx,paralist, amssymb}
\usepackage{stfloats}

\usepackage{balance}  
\usepackage[ruled,noend,noline,vlined,linesnumbered]{algorithm2e}
\usepackage{mathtools,pict2e,epic,relsize}
\usepackage{xfrac}
\usepackage{subfig}
\usepackage{setspace}
\usepackage{scalerel}
\usepackage{natbib}

\newcommand*\btrfly{\DOTSB\mathbin{\mathpalette\btrflyaux\relax}}

\newlength\ztimespadding
\newcommand*\ltaux[2]{\vcenter{\hbox{\hspace{\ztimespadding}%
  \ifx#1\displaystyle
    \setlength\unitlength{1ex}%
    \linethickness{.1ex}%
  \else\ifx#1\textstyle
    \setlength\unitlength{1ex}%
    \linethickness{.1ex}%
  \else\ifx#1\scriptstyle
    \setlength\unitlength{.8ex}%
    \linethickness{.09ex}%
  \else\ifx#1\scriptscriptstyle
    \setlength\unitlength{.65ex}%
    \linethickness{.07ex}%
  \fi\fi\fi\fi
  \begin{picture}(1,1)\roundcap
    \put(0.5,0.5){\circle{1.5}}
    \put(0,0){\relsize{-4}{1}}
  \end{picture}%
  \hspace{\ztimespadding}%
}}}
\newcommand*\rtaux[2]{\vcenter{\hbox{\hspace{\ztimespadding}%
  \ifx#1\displaystyle
    \setlength\unitlength{1ex}%
    \linethickness{.1ex}%
  \else\ifx#1\textstyle
    \setlength\unitlength{1ex}%
    \linethickness{.1ex}%
  \else\ifx#1\scriptstyle
    \setlength\unitlength{.8ex}%
    \linethickness{.09ex}%
  \else\ifx#1\scriptscriptstyle
    \setlength\unitlength{.65ex}%
    \linethickness{.07ex}%
  \fi\fi\fi\fi
  \begin{picture}(1,1)\roundcap
    \put(0.5,0.5){\circle{1.5}}
    \put(0,0){\relsize{-4}{2}}
  \end{picture}%
  \hspace{\ztimespadding}%
}}}

\newcommand*\ctplraux[2]{\vcenter{\hbox{\hspace{\ztimespadding}%
  \ifx#1\displaystyle
    \setlength\unitlength{1ex}%
    \linethickness{.2ex}%
  \else\ifx#1\textstyle
    \setlength\unitlength{1ex}%
    \linethickness{.2ex}%
  \else\ifx#1\scriptstyle
    \setlength\unitlength{.8ex}%
    \linethickness{.18ex}%
  \else\ifx#1\scriptscriptstyle
    \setlength\unitlength{.65ex}%
    \linethickness{.18ex}%
  \fi\fi\fi\fi
  \begin{picture}(1,1)\roundcap
    \put(0,0){\line(1,0){1}}
    \put(1,0){\line(-1,1){1}}
    \put(0,1){\line(1,0){1}}
  \end{picture}%
  \hspace{\ztimespadding}%
}}}

\newcommand*\btrflyaux[2]{\vcenter{\hbox{\hspace{\ztimespadding}%
  \ifx#1\displaystyle
    \setlength\unitlength{1.5ex}%
    \linethickness{.1ex}%
  \else\ifx#1\textstyle
    \setlength\unitlength{1.5ex}%
    \linethickness{.1ex}%
  \else\ifx#1\scriptstyle
    \setlength\unitlength{1.2ex}%
    \linethickness{.1ex}%
  \else\ifx#1\scriptscriptstyle
    \setlength\unitlength{.975ex}%
    \linethickness{.1ex}%
  \fi\fi\fi\fi
  \begin{picture}(1,1)\roundcap
    \put(0,0.27){\line(1,0){1}}
    \put(1,0.27){\line(-1,1){1}}
    \put(0,1.27){\line(1,0){1}}
    \put(0,0.27){\line(1,1){1}}
  \end{picture}%
  \hspace{\ztimespadding}%
      \vspace{-0.3ex}
}}}

\SetKwProg{Fn}{Function}{}{}
\makeatletter
\g@addto@macro\bfseries{\boldmath}
\makeatother

\let\oldnl\nl
\newcommand{\nonl}{\renewcommand{\nl}{\let\nl\oldnl}}

\newcommand{\meach}{\text{\textnormal{\textbf{each} }}}

\DontPrintSemicolon

\newtheorem{defn}{Definition}

\newcommand{\etal}{\emph{et al.}}

\newcommand{\cB}{\mathcal{B}}

\newcommand{\con}{{\tt condmat}}
\newcommand{\db}{{\tt dbconf}}
\newcommand{\git}{{\tt github}}
\newcommand{\mar}{{\tt marvel}}
\newcommand{\imdb}{{\tt IMDb}}
\newcommand{\dblp}{{\tt DBLP}}
\newcommand{\disaff}{{\tt d-label}}
\newcommand{\dissty}{{\tt d-style}}
\newcommand{\itwiki}{{\tt wiki-it}}
\newcommand{\kindle}{{\tt kindle}}

\newcommand{\tip}[1]{{\theta}(#1)}
\newcommand{\wing}[1]{{\psi}(#1)}

\newcommand{\rd}[1]{\textcolor{purple}{#1}}

\begin{document}

\copyrightyear{2018}
\acmYear{2018}
\setcopyright{usgovmixed}
\acmConference[WSDM 2018]{WSDM 2018: The Eleventh ACM International Conference on Web Search and Data Mining }{February 5--9, 2018}{Marina Del Rey, CA, USA}
\acmPrice{15.00}
\acmDOI{10.1145/3159652.3159678}
\acmISBN{978-1-4503-5581-0/18/02}

\title{Peeling Bipartite Networks for Dense Subgraph Discovery}

\author{Ahmet Erdem Sar{\i}y\"{u}ce}
\affiliation{
  \institution{University at Buffalo}
}
\email{erdem@buffalo.edu}
\author{Ali Pinar}
\affiliation{
  \institution{Sandia National Laboratories}
}
\email{apinar@sandia.gov}

\begin{abstract}
Finding dense bipartite subgraphs and detecting the relations among them is an important problem for affiliation networks that arise in a range of domains, such as social network analysis, word-document clustering, the science of science, internet advertising, and bioinformatics.
However, most dense subgraph discovery algorithms are designed for classic, unipartite graphs.
Subsequently, studies on affiliation networks are conducted on the co-occurrence graphs (e.g., co-author and co-purchase) that project the bipartite structure to a unipartite structure by connecting two entities if they share an affiliation.
Despite their convenience, co-occurrence networks come at a cost of loss of information and an explosion in graph sizes, which limit the quality and the efficiency of solutions.
We study the dense subgraph discovery problem on bipartite graphs.
We define a framework of bipartite subgraphs based on the butterfly motif (2,2-biclique) to model the dense regions in a hierarchical structure.
We introduce efficient peeling algorithms to find the dense subgraphs and build relations among them.
We can identify denser structures compared to the state-of-the-art algorithms on co-occurrence graphs in real-world data.
Our analyses on an author-paper network and a user-product network yield interesting subgraphs and hierarchical relations such as the groups of collaborators in the same institution and spammers that give fake ratings.

\end{abstract}

\maketitle

%


\vspace{-3ex}
\section{Introduction}\label{sec:intro}

Many real-world systems are naturally modeled as affiliation, two-mode, or bipartite networks~\cite{Borgatti97, Latapy08}.
In a bipartite network, vertices are decomposed into two disjoint sets, {\em primary} and {\em secondary}, such that edges can only connect vertices from different sets.
For example authors and papers can be the primary and secondary vertex sets, with an edge representing authorship.
Finding dense subgraphs in the real-world affiliation networks, and relating them to each other has been shown to be useful across different domains.
Literature is rich with the examples such as spam group detection in web~\cite{Gibson05}, word and document clustering~\cite{Dhillon01}, and sponsored search advertising on webpages~\cite{Fain06}.
Despite their representation power, bipartite graphs are underutilized, since most graph mining algorithms, including dense subgraph discovery, are studied on the traditional unipartite graphs.
For this reason, affiliation networks are projected to co-occurrence graphs, such that two vertices in the primary set are connected by an edge if  they share an affiliation.
For instance, an author-paper network can be transformed into a co-authorship network, where two authors are connected if they co-authored a paper.
However, this transformation comes at a cost  of information loss and inflated graph size, as we will discuss in more detail later.
Therefore, designing algorithms that can work directly on the bipartite graph, which provides an accurate representation of the underlying system, is essential. 

\begin{figure}[!t]
\centering
\vspace{-1ex}
\captionsetup[subfigure]{labelformat=empty, captionskip=-1.5ex}
\hspace{-8ex}
\subfloat[]{\includegraphics[width=0.47\linewidth]{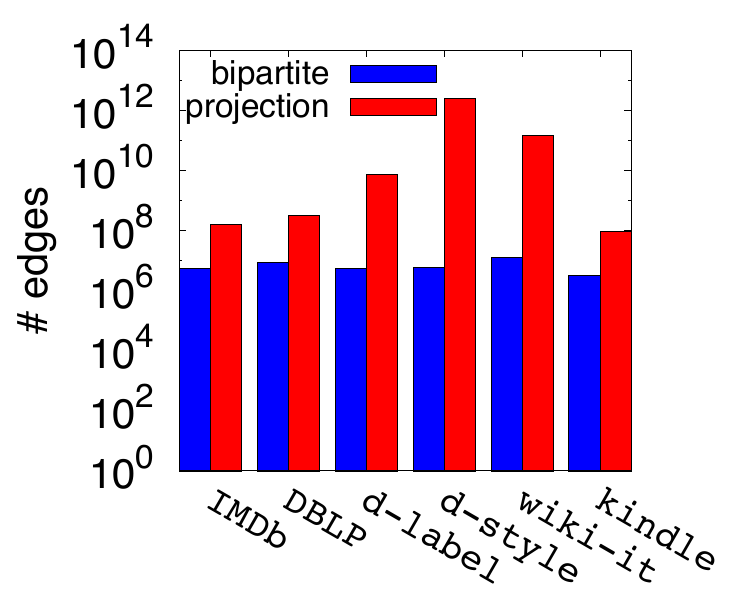}}
\hspace{-2ex}
\subfloat[]{\includegraphics[width=0.47\linewidth]{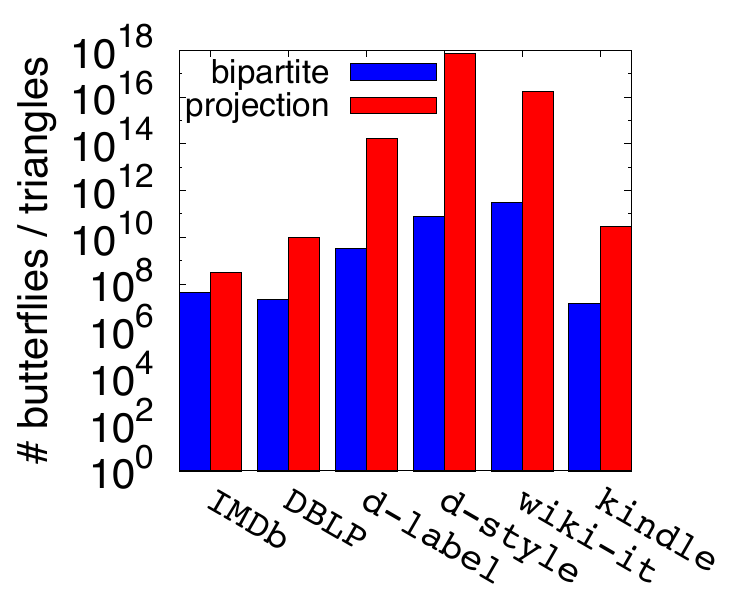}\label{fig:ho}}
\hspace{-12ex}
\vspace{-4.7ex}
\caption{\small Number of edges and higher-order structures in bipartite graphs and their projections}
\label{fig:bars}
\vspace{-1ex}
\end{figure}

This paper studies  finding dense subgraphs in a {\em bipartite} graph and detecting the relations among them.
Our approach is inspired by the $k$-core~\cite{Seidman83,MaBe83} and $k$-truss~\cite{Saito06, Cohen08, Zhang12, Verma12} decompositions in unipartite networks, which are instances of  {\em peeling algorithms}.
They have been shown to be effective to detect dense subgraphs with hierarchical relations~\cite{Sariyuce15}.

\vspace{-2ex}
\subsection{Problem and Challenges}
Our aim is to find many, if not all, dense regions in bipartite graphs and determine the relations among them.
Despite a few successful studies that directly focus on bipartite networks~\cite{Chen12, Beutel13}, a common practice in the literature  for working with bipartite graphs has been creating  co-occurrence (projection)  graphs. 
Although the projection enables the use of well-studied unipartite graph mining algorithms~\cite{Borgatti97}, it has significant drawbacks:

\begin{compactitem}[\leftmargin=-.15in$\bullet$]
\item \textbf{Information loss and ambiguity:} Bipartite graphs comprise one-to-many relationship information, but  this information is  reduced to pairwise ties when projected to a weighted or unweighted unipartite form.
Those pairwise ties are treated independently, which distorts the original information.
In addition, projections are not bijective irrespective of the projection technique being used, which  creates ambiguity.
\item[$\bullet$] \textbf{Size inflation:} A secondary vertex in a bipartite network with degree $d$ results in a $d$-clique in the projected graph.
Thus, the number of edges in the projected graph can be as many as $\sum_{v \in V}{d_v\choose 2}$, whereas it is only $\sum_{v \in V}{d_v}$ in the bipartite network, where $V$ is the set of secondary vertices.
Fig.~\ref{fig:bars} (left) shows the difference between some bipartite networks and their projections -- we observe up to 6 orders of magnitude increase in size, which degrades the performance.
This also artificially boosts the clustering coefficients and the local density measures in the projected graph by creating many triangles. Regarding the smallest higher-order structures, a projection can have up to $699$ quadrillion triangles whereas its original bipartite network has $77$ billion butterflies (2,2-biclique), as in Fig.~\ref{fig:bars} (right).
\end{compactitem}

Given the drawbacks of the projection approaches, we work directly on the bipartite graph to discover the dense structures.
It has been shown that the higher-order structures (motifs, graphlets) offer deeper insights for analyzing real-world networks and detecting dense regions in a better way~\cite{Sariyuce15, Benson16, Babis17,Huang14,Tsourakakis15}.
Peeling algorithms, $k$-core and $k$-truss decompositions, find dense regions in unipartite graphs and determine the relations among them~\cite{Huang14, Zhang12, Gregori11}.
Nucleus decomposition~\cite{Sariyuce15} is a generalization of these two approaches and can work on higher-order structures such as 4-cliques.
However, none of them are applicable for the bipartite networks.
$k$-core decomposition assumes that all vertices represent the same kind of entity, which does not hold for  bipartite graphs.
$k$-truss decomposition works on triangles, which do not exist in bipartite graphs.
Nucleus decomposition uses small-cliques, which are also nonexistent in bipartite graphs.
Thus, we need \textbf{higher-order structures that capture the cohesiveness in bipartite graphs}.
Then the peeling algorithms can be adapted to run on these structures to find the dense regions with hierarchical relations. 

\begin{table}[!t]
\vspace{-2ex}
\small
\renewcommand{\tabcolsep}{2pt}
\linespread{0}\selectfont{}
\caption{\small Notations}
\vspace{-4ex}
\linespread{1}\selectfont{}
\begin{tabular}{ | l | l |} \hline
$G=(U,V,E)$		& bipartite graph with vertices in U and V, and edges E \\ \hline
$N(u,G)$, $N(u)$	& set of vertices that are connected to vertex $u$ in $G$ \\ \hline
$d(u)$			& degree of vertex $u$, i.e., $|N(u)|$ \\ \hline
$(a,b)$-biclique		& complete bipartite graph where $|U|=a$ and $|V|=b$\\ \hline
$G_p=(U, E_p)$	& unweighted projection of $G$, as in Def.~\ref{def:weproj}\\ \hline
$G_{wp}=(U,E_{wp})$& weighted projection of $G$, as in Def.~\ref{def:weproj}\\ \hline
$\triangle_p$		& number of triangles in the projected graph $G_p$ \\ \hline
$\btrfly$			& butterfly or $(2,2)$-biclique \\ \hline
$\tip{u}$			& tip number of vertex $u$, as in Def.~\ref{def:tipno} \\ \hline
$\wing{e}$			& wing number of edge $e$, as in Def.~\ref{def:wingno} \\ \hline
\end{tabular}
\label{tab:notation}
\end{table}

\vspace{-2ex}
\subsection{Contributions}
\begin{figure}[!t]
\centering
\captionsetup[subfigure]{captionskip=-3ex}
\vspace{-4ex}
\hspace{16ex}
\includegraphics[width=0.4\linewidth]{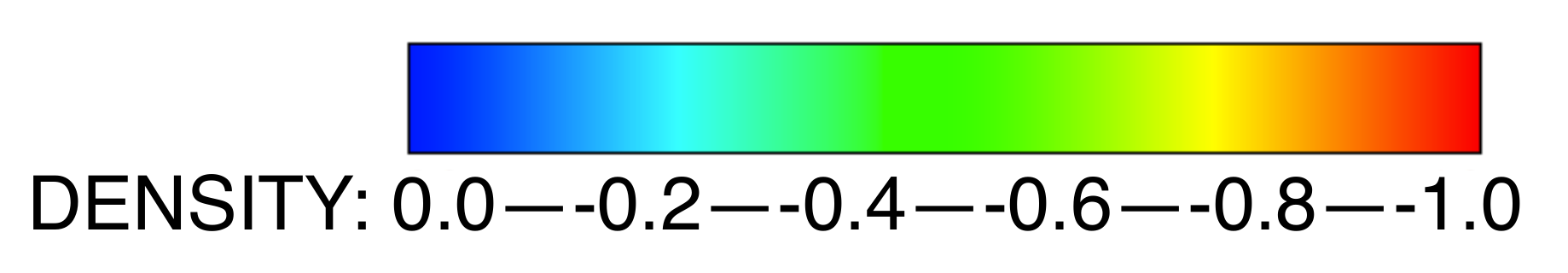}
\vspace{-6ex}
\hspace{-25ex}

\hspace{-12ex}\small
\subfloat[\small \textbf{\textsc{Wing}}]{\includegraphics[width=0.6\linewidth]{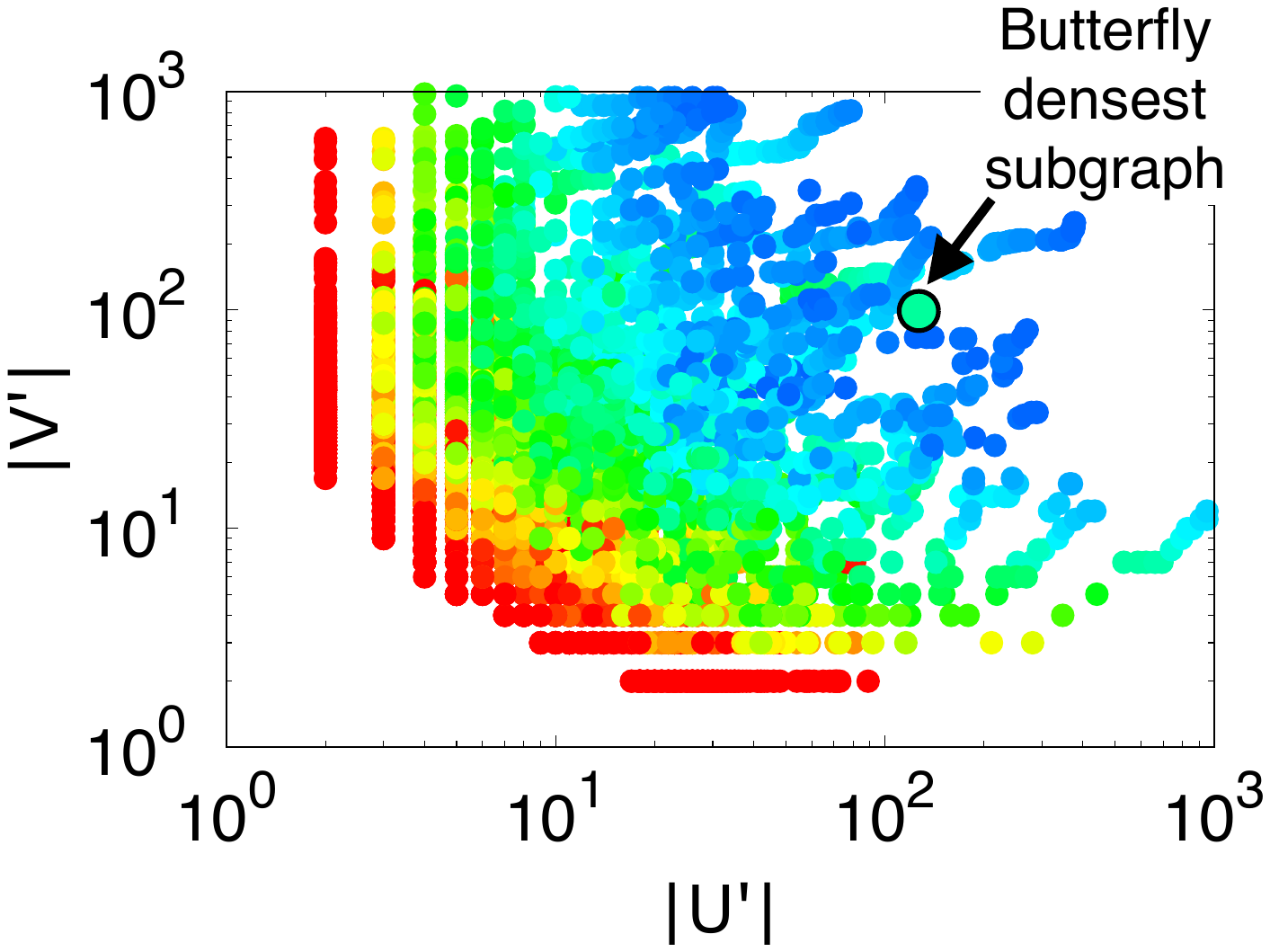}}
\hspace{-5ex}
\subfloat[\small \textbf{\textsc{$(2,3)$ nucleus}}]{\includegraphics[width=0.6\linewidth]{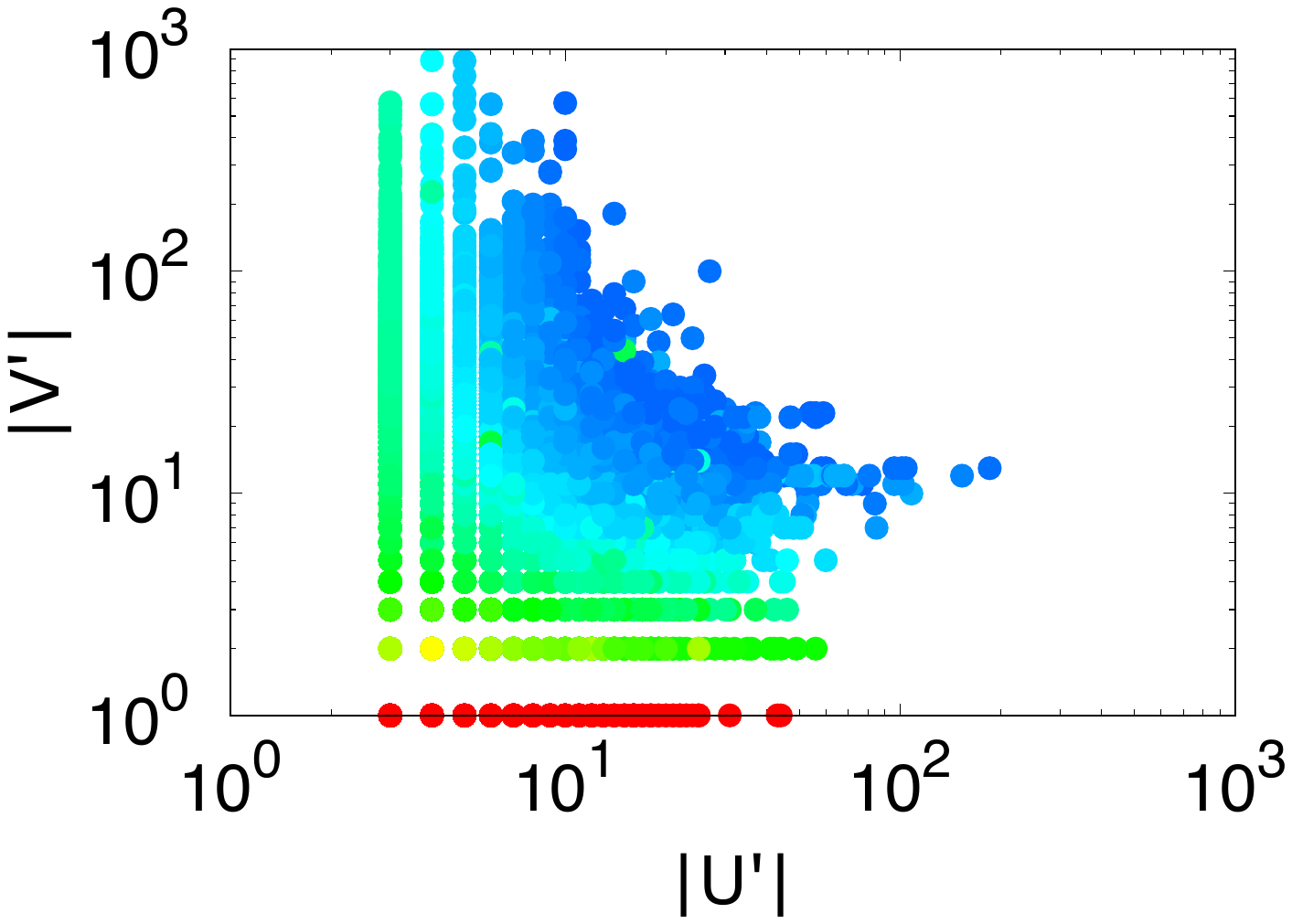}}
\hspace{-13ex}

\hspace{-1ex}
\vspace{-7.5ex}
\caption{\small \bf Dense subgraphs for the \imdb~network. Each dot is a bipartite subgraph and edge density is color coded. $U'$ and $V'$ are the primary and secondary vertex sets and their sizes are given on the x- and y-axes. Wing decomposition results in 36 bipartite subgraphs with $\ge 0.9$ edge density that have at least 10 vertices in each side, and perform competitive to the butterfly densest subgraph~\protect{\cite{Babis15KDD}}. Other algorithms working on projections, on the right, cannot report any bipartite subgraphs in that quality.}
\vspace{0ex}
\label{fig:introfig}
\end{figure}

We introduce new algorithms to efficiently find dense bipartite subgraphs with hierarchy of relations.
Our contributions can be summarized as follows:

\begin{compactitem}[\leftmargin=-0.1in$\bullet$]
\item \textbf{Introducing $k$-tip and $k$-wing bipartite subgraphs:} We survey  attempts to define higher-order structures in bipartite graphs, and use the {\em butterfly} structure (2,2-biclique) as the simplest super-edge motif.
Building on that, we define the {\em $k$-tip} and {\em $k$-wing} subgraphs based on the involvements of vertices and edges in butterflies, respectively.
\item \textbf{Extension  of peeling algorithms:} We introduce peeling algorithms to efficiently find all the $k$-tip and $k$-wing subgraphs. 
Our algorithms are inspired by the degeneracy based decompositions for unipartite graphs.
We present detailed psueducodes and analyze their complexities.
\item \textbf{Evaluation on real-world data:} We evaluate our proposed techniques on real-world networks.
Fig.~\ref{fig:introfig} presents a glance of results on the IMDb movie-actor with $1.6$M vertices and $5.6$M edges.
Our algorithms are able to extract  larger and denser subgraphs of various sizes.
We also examine the ratings data for the Amazon Kindle books and analyze the author-paper network of the top database conferences.
We highlight the interesting subgraphs and hierarchies we detect that cannot be discovered by the existing approaches.
Finally, we present the runtime performances.
\end{compactitem}

\vspace{-2ex}
\section{Background}

This section reviews basics about the bipartite networks and the peeling algorithms.
We present our notation in Table~\ref{tab:notation}.

Let $G=(U,V,E)$ be an undirected, unweighted, simple (no loop, no multi-edge) bipartite graph.
$U$ is the set of primary vertices, $V$ is the set of secondary vertices, and $E$ is the set of edges s.t. $\forall (u, v) \in E$, $u \in U \land v \in V$.
$N(u, G)$ denotes the neighbor set of a vertex $u $ in the bipartite graph.
We abuse the notation by using $N(u)$ when $G$ is obvious.
$d(u)$ is the degree of the vertex $u$ and defined as $|N(u)|$.
We define the density of a bipartite subgraph $G=(U, V, E)$ as the ratio of the number of existing edges over the number of all possible edges, i.e., $\frac{|E|}{|U|\cdot|V|}$.
$H=(U',V',E')$ is an \textbf{induced subgraph} of the bipartite graph $G=(U,V,E)$, if~$U' \subseteq U$, $V' \subseteq V$, and
$V'=\cup_{u \in U'}{N(u, G)}$, $E'=\cup_{u \in U'}{\cup_{v \in V'}{(u, v)}}$. $G=(U,V,E)$ is an \textbf{(a,b)\textnormal{-}biclique} if it is a complete graph between $a$ vertices on one side and $b$ vertices on the other.

\begin{figure}[!t]
\centering
\vspace{-1ex}
\hspace{-40ex}
\begin{minipage}{3.5cm}
\captionsetup[subfigure]{font={stretch=0.7}, captionskip=1ex}
\subfloat[\small \bf The bipartite graph on the left is projected to the weighted unipartite graph on the right. $A$, $B$, and $C$ form a triangle since they are all affiliated with the same vertex. Vertex $D$ only connects to $C$ in the projection since it is the only one with which it shares an affiliation.
The edge between vertices $A$ and $B$ is assigned $1/3 + 1/2$, because one of the affiliations they share in the bipartite graph has 3 neighbors and the other affiliation has 2 neighbors.]{\includegraphics[width=3.5cm, keepaspectratio]{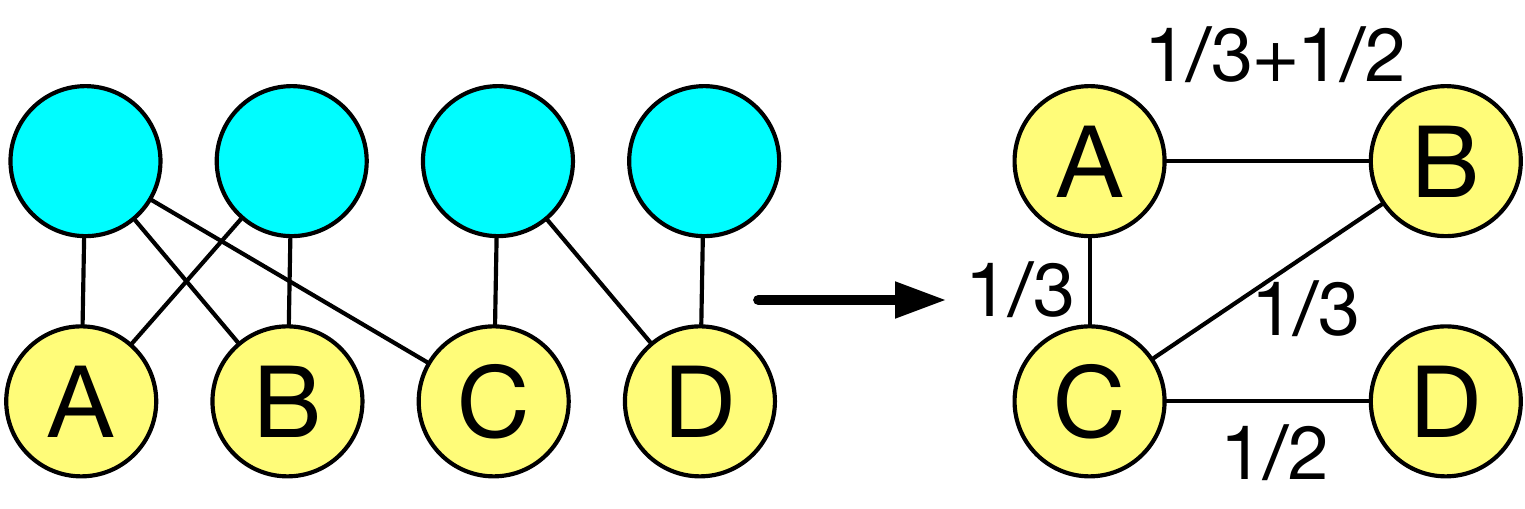}\label{fig:exproj}}
\vspace{-1ex}

\subfloat[\small \bf Graph motifs to model cohesion in bipartite networks.]{\hfill\includegraphics[width=1\textwidth, keepaspectratio]{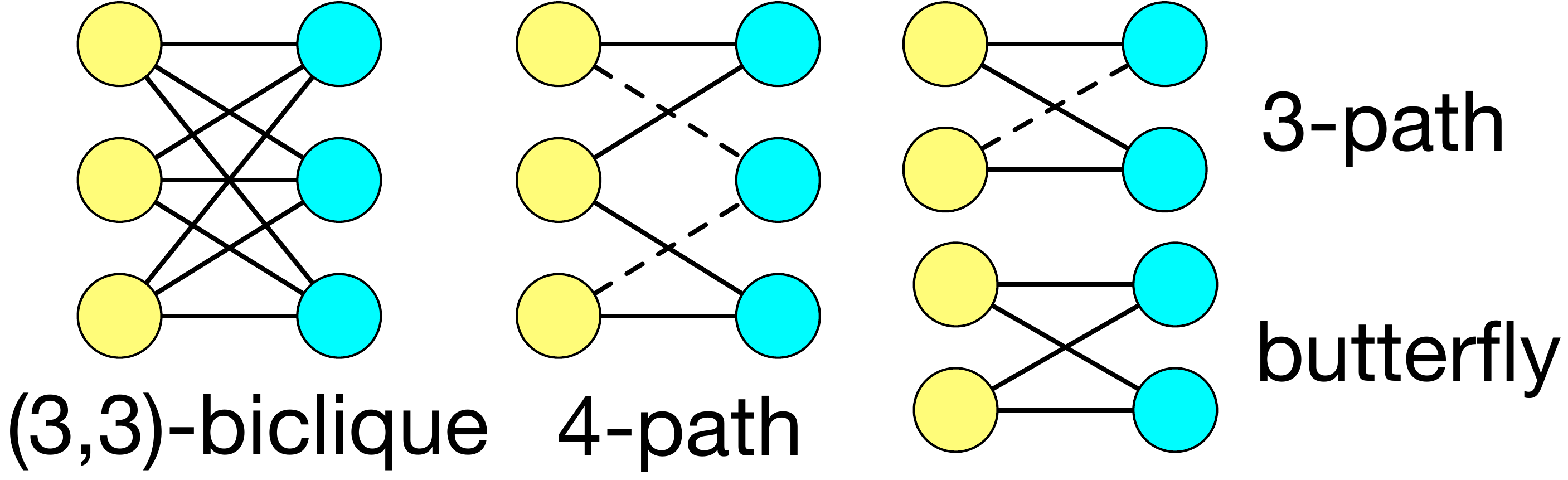}\label{fig:coh}}
\end{minipage}
\hspace{1ex}
\vspace{-5ex}
\begin{minipage}{4.5cm}
\captionsetup[subfigure]{font={stretch=0.7}, captionskip=-0.7ex}
\subfloat[\small \bf The entire graph is a $2$-core since each vertex have $\ge$ degree 2. Each triangle is a $1$-truss, denoted in dashed lines, since each edge takes part in one triangle. Two $1$-trusses are not merged, because the edge in the middle has no triangle.]{\label{fig:nuc1}\includegraphics[width=4.5cm, keepaspectratio]{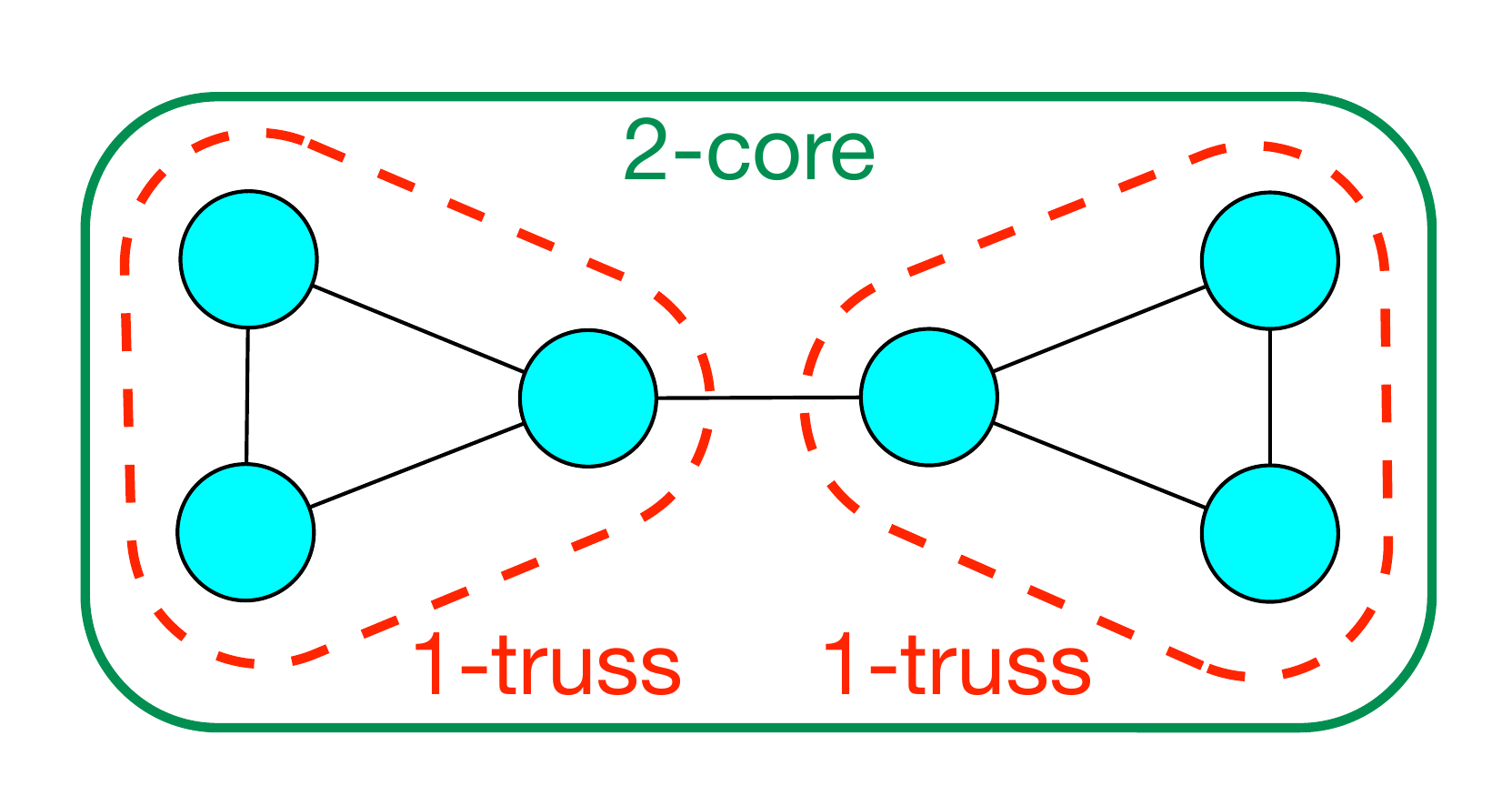}}
\vspace{-2ex}

\captionsetup[subfigure]{font={stretch=0.7}, captionskip=-0.3ex}
\subfloat[\small \bf The entire graph is a $1$-truss since each edge has 1 triangle. There are two $1$-(2,3) nucleus subgraphs, overlapping on the middle vertex. These two nuclei are not merged, because no triangle exists that contains an edge from each nucleus.]{\label{fig:nuc2}\includegraphics[width=4.5cm, keepaspectratio]{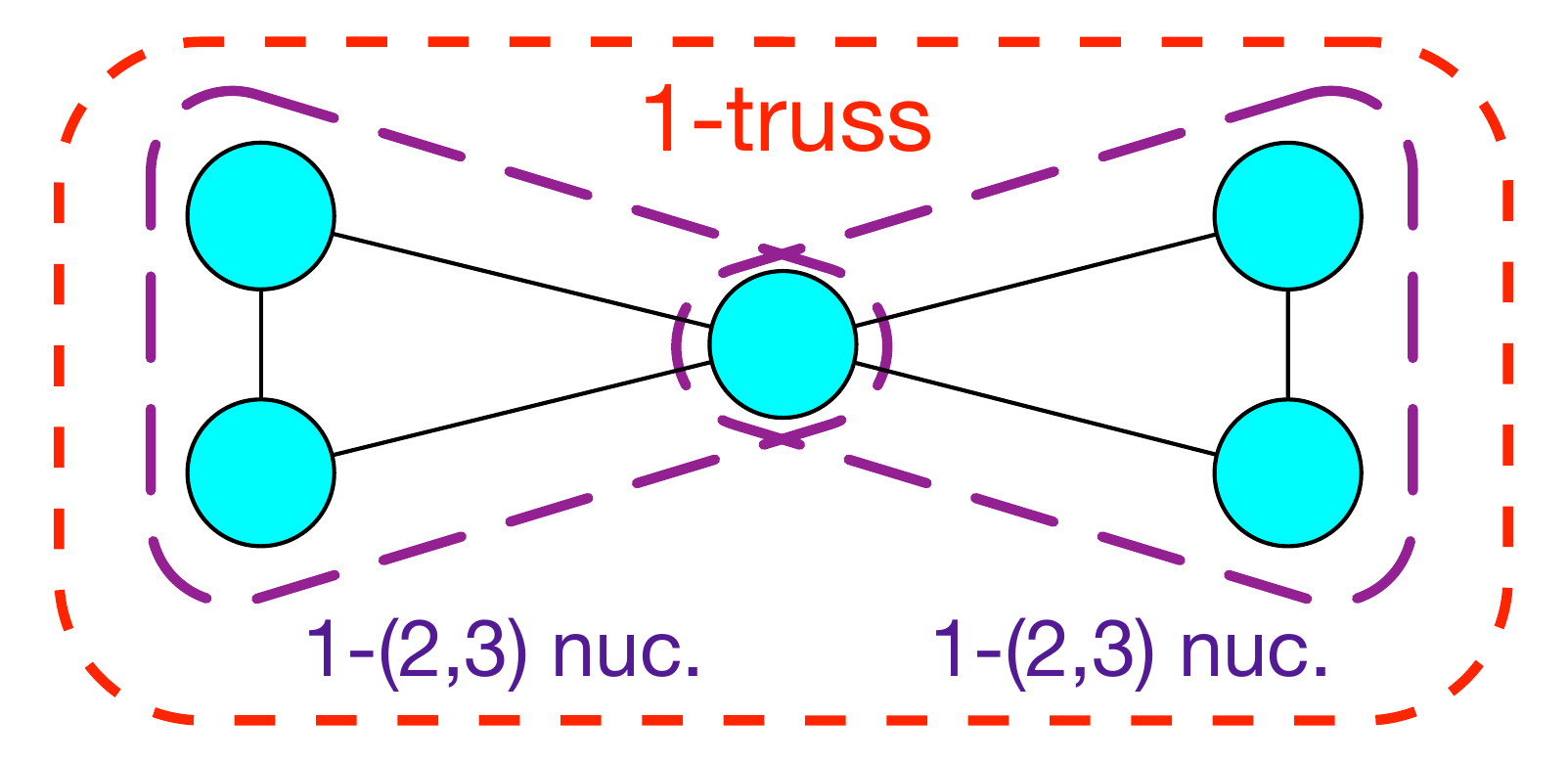}}
\end{minipage}
\hspace{-40ex}
\vspace{2ex}
\caption{\small \bf Examples for projections, and $k$-core, $k$-truss and $k$-(2,3) nucleus}
\end{figure}

We present  two ways to convert a bipartite graph to a unipartite graph~\cite{Newman01a, Newman01b}, also illustrated in Fig.~\ref{fig:exproj}.
Weighted projection is built by assigning weights to the edges.
Weights are computed in proportion to the number of vertices connected to each affiliation in the bipartite graph.

\begin{defn} \label{def:weproj} 
Given a bipartite graph $G=(U, V, E)$, its \textbf{weighted projection} is an edge-weighted unipartite graph $G_{wp}=(V_{wp}, E_{wp})$ s.t. $V_{wp}=U, E_{wp}=\{(u_1,u_2, w)~|~N(u_1) \cap N(u_2)~\neq \varnothing \land w=\sum_{v \in (N(u_1) \cap N(u_2))}\frac{1}{|N(v)|}\}$.
\textbf{Unweigted projection} is the same as weighted projection where the edge weights are $1$.
\end{defn}

$k$-core~\cite{Seidman83, MaBe83} and $k$-truss~\cite{Saito06, Cohen08, Zhang12, Verma12} subgraphs, which inspire our approach, are defined as follows:

\begin{defn}\label{def:kcore}
A connected subgraph, $H$, of $G$ is a \textbf{k-core} if every vertex in $H$ has at least degree $k$ and no other subgraph of $G$ that subsumes $H$ is a $k$-core.
\textbf{Core number} of a vertex $u$ is the maximum $k$ such that  there is a $k$-core subgraph that contains $u$.
\end{defn}

To find  $k$-cores, vertices with degree  $< k$ and their edges  are removed from the graph, until no such vertex remains.
For a full decomposition, we increment $k$ at each step, and assign $k$ as  the core number the removed vertex. This process is called as `peeling', and it works in $O(|E|)$ time~\cite{BaZa03}.

\begin{defn}\label{def:ktruss}
A connected subgraph, $H$, of $G$ is a \textbf{k-truss} if each edge in $H$ takes part in $\geq k$ triangles and  no other subgraph of $G$ that subsumes $H$ is a $k$-truss.
\end{defn}

Nucleus decomposition is a generalization of $k$-core and $k$-truss decompositions.
Instead of vertex-edge or edge-triangle relations, nucleus decomposition works on any clique relations.
The idea is described in \cite{Sariyuce15}, but we restrict presentation to a specific case for brevity.
Here we only define $k$-(2,3) nucleus to highlight its stronger connectedness than the $k$-truss.
It is also referred to as `$k$-truss community' in~\cite{Huang14}.

\begin{defn}\label{def:23}
\hspace*{-1.5ex}
A subgraph $H\!=\!\!(V, \!E)$  of $G$ is a \textbf{k-(2,3)-nucleus}, iff
\begin{compactitem}[\leftmargin=-1.5ex $\bullet$]
\item  each edge takes part in at least $k$ triangles;
\item any pair of edges in $E$ is connected by series of triangles; 
\item no other subgraph of $G$ that subsumes $H$ is a  \textbf{k-(2,3)-nucleus}.
\end{compactitem}
\end{defn}

Here $(2,3)$ refers to the 2-clique (edge) and 3-clique (triangle) relations.
Two edges $e$ and $f$ are connected by \textit{series of triangles} if there exists a sequence of edges $e = e_1, e_2, \ldots, e_k = f$ such that for each $i$, some triangle contains $e_i$ and $e_{i+1}$.
In Fig.~\ref{fig:nuc1}, entire graph is a 2-core.
Two separate 1-trusses appear since the middle edge has no triangle.
In Fig.~\ref{fig:nuc2}, each edge takes part in 1 triangle, making the entire graph $1$-truss.
However, two separate $1$-($2,3$) nuclei exist, because there is no triangle that connect edges from each nucleus.

\vspace{-1ex}
\section{Related Work}\label{sec:rel}

Literature on bipartite network analysis has two main thrusts:  extending unipartite graph concepts to bipartite graphs and defining new projection methods to get unipartite representations.
Centrality and the density metrics~\cite{Borgatti97}, clustering coefficients~\cite{Robins04}, matrix partitioning~\cite{Umit99} and clustering~\cite{Wolf16} algorithms are adapted for bipartite graphs.
Regarding the projections, Newman introduced the weighted projection for scientific collaboration networks~\cite{Newman01a, Newman01b} and Everett and Borgatti proposed to use dual projections ~\cite{Everett13}.
We focus on using the actual bipartite graph (without projection) to find many dense subgraphs and their relations to each other, which have not been explored thoroughly.

\begin{figure*}[!t]
\vspace{-3ex}
\centering
\captionsetup[subfigure]{captionskip=-0.4ex}
\vspace{-4ex}
\hspace{-8ex}
\subfloat[\small \bf We check the vertices on the bottom to find $k$-tips.
$g$ has no butterfly.
Vertices $a, b, e,$ and $f$ take part in two butterflies while $c$ and $d$ are involved in three.
But $c$ and $d$ cannot have a tip number of $3$ since their induced subgraph has just one butterfly.
Thus, vertices $a$ to $f$ forms a $2$-tip.
Each vertex gets the largest $k$-tip that they are part of; $\tip{a-f}$=$2$.]
{
\includegraphics[width=0.35\linewidth]{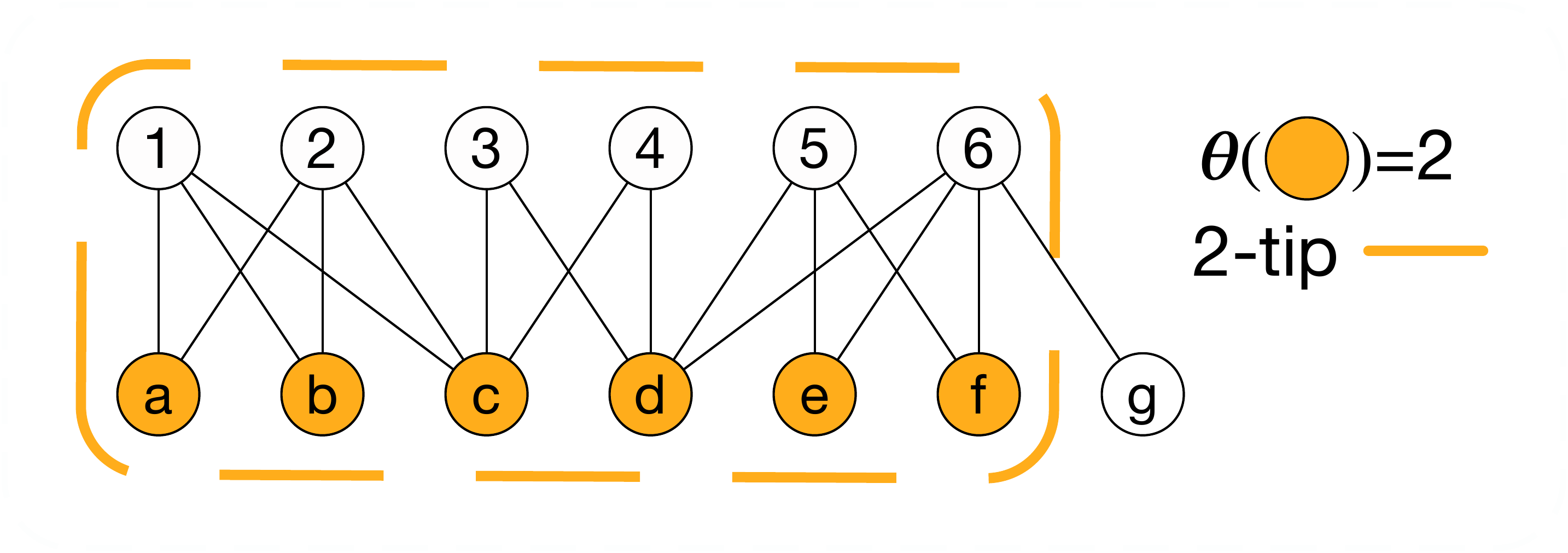}
\label{fig:tip}}
\hspace{1ex}
\subfloat[\small \bf $k$-wings on the same graph.
Edge $(g,6)$, $g6$ in short, has no butterfly.
Each of the four edges in the middle, $c3, c4, d3, d4$, participate in only one butterfly, thus each edge has a wing number of $1$ and they form a $1$-wing.
There are also two $(3,2)$-bicliques; $abc12$ and $def56$.
Each edge in those bicliques takes part in two butterflies.
So, each is a $2$-wing, and all the edges in those have a wing number of $2$.
Overall, $k$-wings can reveal denser regions than the $k$-tips on the given toy graph.]
{\includegraphics[width=0.35\linewidth]{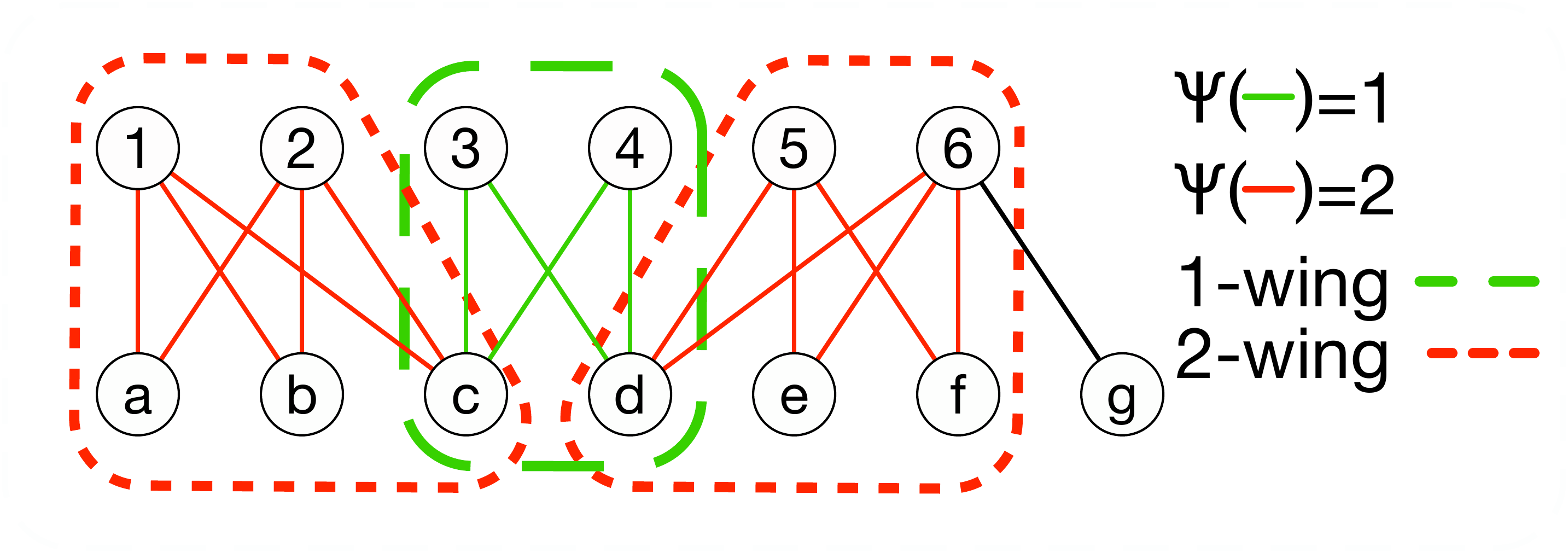}
\label{fig:wing}}
\hspace{1ex}
\captionsetup[subfigure]{captionskip=-0.4ex}
\subfloat[\small \bf In an author-paper network, the author shown in red cannot be considered in a single research community because she collaborates with different researchers on distinct sets of papers. Each affiliation of the author should be considered independently to better detect the communities she is involved in.]
{\includegraphics[width=0.2\linewidth]{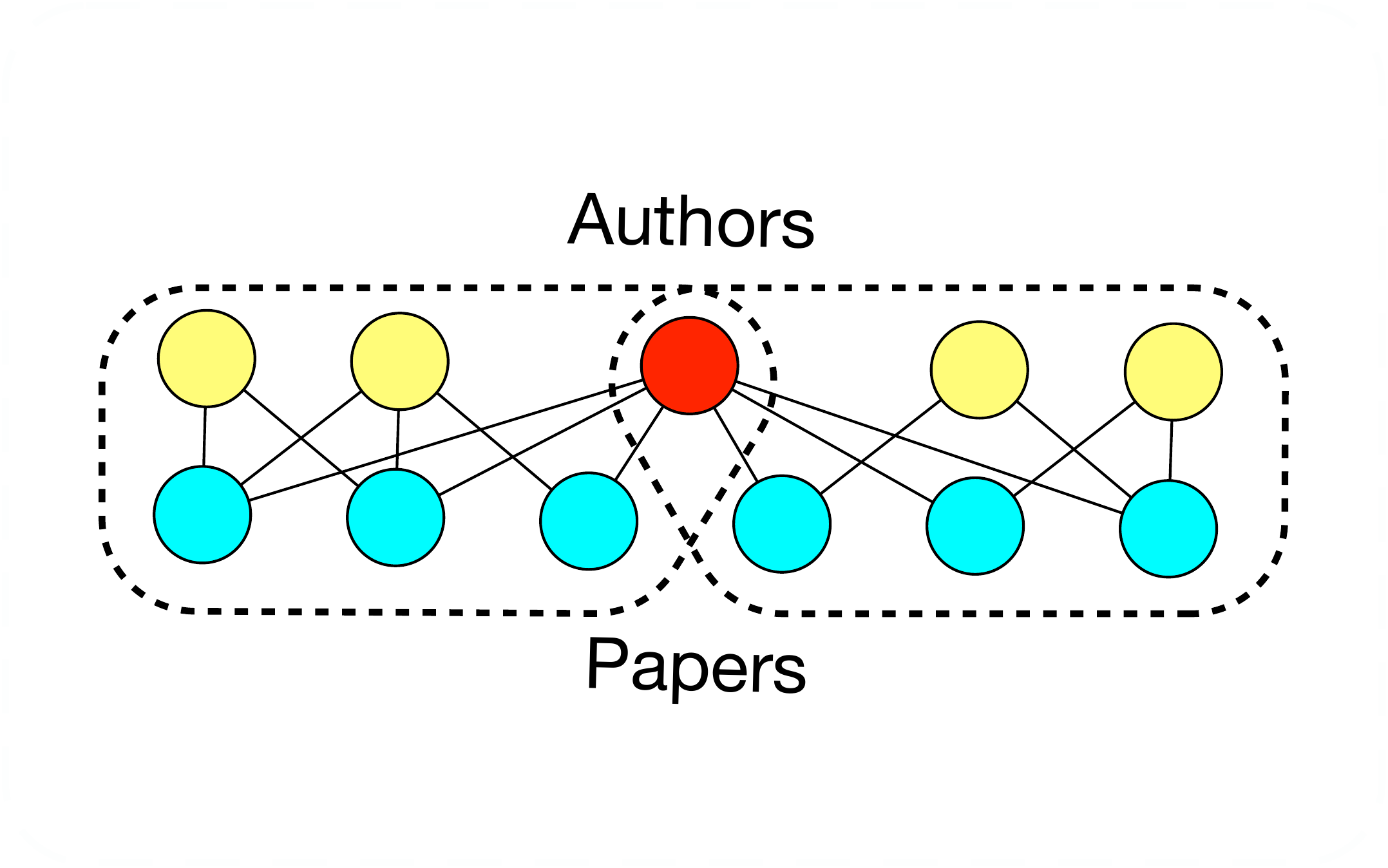}
\label{fig:strongconn}}
\hspace{-7ex}
\vspace{-2ex}
\caption{\small \bf Illustrations of the $k$-tip and $k$-wing subgraphs, and tip ($\theta$) and wing ($\psi$) numbers.}
\vspace{-3ex}
\end{figure*}

\textbf{Bipartite graph motifs}: Capturing the smallest unit of cohesion provides a structural way to find the dense regions.
Various locality patterns~\cite{Borgatti97, Robins04, Opsahl13} and density measures~\cite{Latapy08} have been proposed for bipartite graphs (see Fig.~\ref{fig:coh}).
Borgatti and Everett considered the \textbf{$(3,3)$-biclique} to define cohesiveness~\cite{Borgatti97}.
Opsahl proposed the \textbf{closed $4$-path}, which is also a $(3,3)$-biclique~\cite{Opsahl13}.
Robins and Alexander used the ($2,2$)-biclique to model the cohesion~\cite{Robins04}, and looked for $3$-paths, which consists of three edges with two vertices from each set.
This approach is also adopted in a recent work by Aksoy \etal~\cite{Aksoy16} to generate bipartite graphs with community structure, where ($2,2$)-biclique is called a \textbf{butterfly}.

\textbf{Bipartite dense subgraphs}: Fake likes, ratings and reviews are prevalent in online social networks and can be modeled by the dense regions in the underlying bipartite network~\cite{Beutel13}.
Regarding quantifying the dense regions, Borgatti and Everett proposed the biclique as a dense subgraph~\cite{Borgatti97}. Kumar \etal~used bicliques of various sizes to analyze web graphs~\cite{Kumar99}.
Enumerating all the maximal bicliques and quasi-cliques is studied by Sim \etal~\cite{Sim09},
and Mukherjee and Tirthapura~\cite{Mukherjee14}.
However, biclique definition is regarded as too strict, not tolerating even a single missing edge, and is also expensive to compute.
More recently, Tsourakakis \etal~\cite{Babis15KDD} used sampling to find $(P, Q)$-biclique densest subgraph in bipartite networks.
Their algorithm results in a single subgraph that has the most number of $(P, Q)$-bicliques in the entire network.
Butterfly ($(2,2)$-biclique) densest subgraph serves as a baseline and we check how our results compare in Sec.~\ref{sec:exps}.
However, we do not focus on finding a single, densest subgraph, but aim to find many dense subgraphs with hierarchical relations.

\textbf{Peeling bipartite networks}: There have been some attempts to adapt peeling algorithms or $k$-core~\cite{Seidman83} like subgraphs to bipartite graphs.
Cerinsek and Batagelj~\cite{Cerinsek15}  adapted the generalized core idea~\cite{Batagelj02} to bipartite networks.
However, their definition is not suitable to construct a hierarchy since it is not clear how to define a comparison function for $(p, q)$ pairs.
Giatsidis \etal~\cite{Giatsidis11} worked on the scientific collaboration networks to find dense regions.
They used the weighted projections and adapted the $k$-core decomposition for weighted networks to detect the hierarchy.
They defined the \textbf{fractional $k$-core} as a maximal subgraph where every vertex has at least weight $k$ (vertex weight is the sum of edge weights that are connected).
Li \etal~\cite{Li13} introduced a $k$-truss like definition for bipartite networks.
They insert artificial edges between vertex pairs that share a neighbor and apply the peeling algorithm on those artificial edges and their triangle counts.
This is actually identical to creating the projection, and applying the $k$-truss decomposition using the triangle counts~\cite{Cohen08}.
In addition, Chen and Saad~\cite{Chen12} proposed algorithms to find many dense structures in bipartite graphs by constructing a dendrogram of the vertices, where the subgraphs are not overlapping.
\textsl{Unlike the previous work, we introduce peeling algorithms that work directly on the bipartite network and identify possible overlapping dense regions in a hierarchy.}

\vspace{-2ex}
\section{Dense bipartite subgraphs}

In many real-world networks, the underlying structure that yields cohesive and close-knit subgraphs is the triangle.
Triangle is the smallest unit of cohesion in uni-partite networks and butterfly can be considered as the \textbf{triangle of bipartite networks} in the sense that it is the smallest cohesive bipartite unit.
Conversely, a bipartite graph does not have a community structure without butterflies.
\textit{Motivated by that, we use the butterfly as the main higher-order structure in our bipartite subgraph models.}
It is the smallest structure with multiple vertices at each side, and also cheaper to enumerate than the larger bicliques.
Our aim is to discover the bipartite subgraphs with many butterflies and construct relations among them.
$k$-core and $k$-truss decompositions reveal the hierarchical relations among dense regions and we follow a similar methodology in our models.
We introduce two bipartite dense subgraph models that have different trade-offs between the subgraph density and the computation time to find.

\subsection{Tip decomposition: the Definition}

We introduce the $k$-tip to identify the vertex-induced subgraphs with many butterflies.
Our  approach enables building hierarchical relations among the subgraphs which results in a global tree structure that represents the significantly dense regions in the graph with various sizes and densities.
$k$-tip measures the intensity of vertex participations in the butterfly structures.
It is defined as follows:

\vspace{-1ex}
\begin{defn} \label{def:tipsubgraph}
A bipartite subgraph $H=(U,V,E) \subseteq G$, induced on $U$, is a \textbf{$k$-tip} iff
\begin{itemize}[\leftmargin=0ex$\bullet$]
\item each vertex $u \in U$ takes part in at least $k$ butterflies,
\item each vertex pair $(u, v) \in U$ is connected by series of butterflies,
\item $H$ is maximal, i.e., there is no other $k$-tip that subsumes $H$.
\end{itemize}
\end{defn}

Two vertices $a$ and $b \in U$ are connected by a \textit{series of butterflies} if there exists a sequence of vertices $a = u_1, u_2, \ldots, u_k = b$ such that some butterfly contains $u_i $ and  $u_{i+1}$, for each $i$.
This connectivity condition helps to separate the dense regions that are only connected by an edge, which are otherwise considered as a single combined subgraph.
Butterfly density of a $k$-tip is obtained by ensuring a lower bound on the number of butterflies that each vertex participates.
Note that a vertex with many butterflies does not imply a dense region by itself; it should also be surrounded by other vertices incident to many butterflies.

Fig.~\ref{fig:tip} illustrates $k$-tip examples, where  $U$  and $V$ cover the vertices on the bottom ($a-g$) and top ($1-6$), respectively.
There are seven butterflies in total: $ab12, ac12, bc12, cd34, de56, df56,$ and $ef56$. Vertex $g$ has no butterflies.
Vertices $c$ and $d$ participate in three butterflies while the others in $U$ (but $g$) are involved in two.
Thus, vertices $a-f$ forms a $2$-tip, shown with the dashed orange line.
$c$ and $d$ cannot form a $3$-tip by themselves or including any other vertex.

Storing all the $k$-tips is not convenient since they are in a hierarchy and have full overlaps.
Instead, we define the \textbf{tip number} of a vertex, denoted as $\theta$, in a similar spirit to the core number (Def.~\ref{def:kcore}), to find any $k$-tip with a traversal operation.

\begin{defn} \label{def:tipno}
\textbf{Tip number}, \textbf{$\tip{u}$}, of vertex $u$, is  the largest {\boldmath $t$} such that  there exists a $t$-tip that contains $u$. \textbf{Tip decomposition} of a graph $G=(U, V, E)$ is finding the tip numbers of vertices in $U$.
\end{defn}

Going back to Fig.~\ref{fig:tip}, we see that all the vertices except $g$ has tip number of $2$ (in orange).
A $k$-tip is found by starting at a vertex with the tip number of $k$ and doing BFS-like traversal on set $U$. At each step, a vertex that resides in a common butterfly is visited if its tip number is at least $k$. Traversal continues until no such vertex appears. Set of visited vertices forms the $k$-tip.

\vspace{-2ex}
\subsection{Wing decomposition: the Definition}

Consider the bipartite network of authors and their papers.
When an author collaborates with different groups on different topics (as in Fig.~\ref{fig:strongconn}),
$k$-tip cannot handle the situation since the dense regions identified by the $k$-tip are defined to be disjoint.
To allow the vertex overlaps, which better captures the dense regions, we introduce another generic subgraph model, \textbf{$k$-wing}.
Its definition is similar to the $k$-tip with one subtle difference: focus is on the {\em edges}, not vertices.
Distinguishing the edges that are connected to the same vertex is the key to get overlaps on the vertices.
We define the \textbf{$k$-wing} bipartite subgraph as follows:

\vspace{-1ex}
\begin{defn}\label{def:wingsubgraph}
A bipartite subgraph $H=(U,V,E) \subseteq G$ is a \textbf{$k$-wing} if,
\begin{itemize}[\leftmargin=0ex$\bullet$]
\item every edge $(u, v) \in E$ takes part in at least $\textbf{k}$ butterflies,
\item each edge pair $(u_1, v_1), (u_2, v_2) \in E$ is connected by series of butterflies,
\item $H$ is maximal, i.e., there is no other $k$-wing that subsumes $H$.
\end{itemize}
\end{defn}

Again we use the butterfly connectedness condition that is stronger than the traditional connectedness and helps to distinguish different regions that are traditionally connected, as shown in Fig.~\ref{fig:strongconn}.
Two edges $a$ and $b \in E$ are connected by a \textit{series of butterflies} if there exists a sequence of edges $a = e_1, e_2, \ldots, e_k = b$ such that some butterfly contains $e_i$ and $e_{i+1}$ for each $i$.

Fig.~\ref{fig:wing} presents the $k$-wings in the same toy graph that we checked in the previous section.
Edge $(g, 6)$, $g6$ in short, does not participate in any butterfly.
Each of the four edges in the middle, $c3, c4, d3, d4$, participate in only one butterfly, thus each edge has a wing number of $1$ and they form a $1$-wing, marked by the green region.
There are also two $(3,2)$ bicliques: $abc12$ and $def56$.
Each edge in those subgraphs participates in exactly two butterflies.
So, each is a $2$-wing, shown in dashed red lines.
Overall, $k$-wings can extract denser regions that cannot be seen by the $k$-tips.
Similar to the tip numbers, we associate each edge with a number to index the subgraphs.
We define the \textbf{wing number} of an edge, denoted by $\psi$, as follows.

\begin{algorithm}[!t]
\small
\caption{\textsc{Tip Decomposition ($G=(U, V, E)$)}}
\label{alg:tipdecom}
\nonl 
\nonl \textcolor{black}{
\Fn{\textbf{{\tt \textbf{D2}} $(u, G)$}}{
\nonl	\textbf{$D \leftarrow \mathlarger{\bigcup}\limits_{v \in N(u)}{\left(\mathlarger{\bigcup}_{d \in N(v)}{\{d\}}\right)}$~\tcp*{\textbf{combine dist-2 neigs}}}
\nonl	\textbf{$c \leftarrow$ multiplicities of $d \in D$}~\tcp*{\textbf{counts of unique items}}
\nonl	\Return \textbf{$c, D$}}
\nonl
\nonl \textcolor{black}{\tcp{find the number of $\btrfly$ each vertex $u \in U$ participates}}
$D \leftarrow$ list, $L \leftarrow$ list, $\beta(\cdot) \leftarrow$ 0, butterfly counts $\forall~u \in U$\;\label{ln:t1}
\For{\meach $u \in U$}{\label{ln:t2}
	$c, D \leftarrow$ \textcolor{black}{{\tt \textbf{D2}} ($u, G$)}\textcolor{black}{~\tcp*{dist-2 neigs ($D$), multiplicity ($c$)}}\label{ln:t3}
	$\beta(u) \leftarrow \sum_{d \in D}{c_d \choose 2}$\textcolor{black}{~\tcp*{update $u$}}\label{ln:t4}
}
\nonl \textcolor{black}{\tcp{find the tip numbers of vertices $u \in U$ by peeling}}
$\tip{u} \leftarrow$ tip numbers $\forall~u \in U$\label{ln:t5}\;
\For{\meach $u$~\textnormal{with minimum} $\beta(u)$~\textnormal{s.t.} $\tip{u}$~\textnormal{is unassigned}}{\label{ln:t6}
  	$\tip{u} \leftarrow \beta(u)$\textcolor{black}{\tcp*{assign the tip number}}\label{ln:t7}
	$c, D \leftarrow$ \textcolor{black}{{\tt \textbf{D2}} ($u, G$)}\textcolor{black}{~\tcp*{dist-2 neigs ($D$), multiplicity ($c$)}}\label{ln:t8}
        \For{\meach $d \in D$~\textnormal{s.t.}~$\tip{d}$~\textnormal{is unassigned}}{\label{ln:t9}
\nonl		\textcolor{black}{\tcp{decrease by the num. of common $\btrfly$, if possible}}
        		\lIf{$\beta(d)-{c_d \choose 2} < \beta(u)$}{$\beta(d) \leftarrow \beta(u)$}\label{ln:t10}
		\lElse {$\beta(d) \leftarrow \beta(d) - {c_d \choose 2}$}\label{ln:t11}
        }
}
\Return array $\tip{\cdot}$\;\label{ln:t12}
}
\end{algorithm}

\begin{defn}\label{def:wingno}
\textbf{Wing number}, \textbf{$\wing{e}$}, of an edge $e \in E$ is the largest {\boldmath $w$} such that there is a $w$-wing that contains $e$. \textbf{Wing decomposition} is the problem of finding the wing numbers of edges in $G$.
\end{defn}
\vspace{-1ex}

In Fig.~\ref{fig:wing}, wing numbers of the edges are shown by the corresponding colors.
$\wing{c3}=\wing{c4}=\wing{d3}=\wing{d4}=1$, the green edges, which belong to the $1$-wing, but not the $2$-wing.
All the other edges, except $g6$, has wing number of $2$, in red.
A $k$-wing can be found by a similar traversal operation explained for the $k$-tip. This time we traverse on edges. We start at an edge with the wing number of $k$ and at each step we visit an edge that resides in a common butterfly and has a wing number of at least $k$. Traversal continues until no such edge remains and the set of visited edges forms the $k$-tip.

\vspace{-1ex}
\section{Peeling Butterflies}\label{sec:alg}

In this section, we present algorithms to find the tip numbers of vertices, $\tip{\cdot}$, and the wing numbers of edges, $\wing{\cdot}$.
$k$-tips and $k$-wings with the hierarchical relations can be located by the $\tip{\cdot}$ and  $\wing{\cdot}$ values, as explained above.

\vspace{-2ex}

\subsection{Tip decomposition: the Algorithm}

We start by counting the number of butterflies that each vertex participates in, and then apply the iterative peeling process where the tip numbers are assigned in a non-decreasing order.
At each step, we find the vertex that has the minimum number of butterflies, assign its current butterfly count as the tip number, say $k$, and decrement the butterfly counts of the vertices (if $>k$) in $U$ that has a common butterfly.

\textsc{Tip Decomposition}, presented in Algorithm~\ref{alg:tipdecom}, finds the tip numbers of the vertices in $U$.
To locate the $k$-tips and construct the hierarchy, we use the disjoint-set forest heuristic that is introduced in a recent work~\cite{Sariyuce17}.
We construct the subgraphs after finding the tip numbers of vertices, and adaptation of~\cite{Sariyuce17} for the tip decomposition is straightforward -- we do not give the details for brevity.

Algorithm~\ref{alg:tipdecom} has two phases.
First, we determine the number of butterflies that each $u \in U$ participates, in lines~\ref{ln:t1} to~\ref{ln:t4}.
A simple way to count the butterflies that a vertex participates is to collect its distance-2 neighbors in a multiset by using a hashmap.
If a vertex $d$ appears $c_d>1$ times in the multiset $D$ (except $u$ itself), then $u$ and $d$ have $c$ common neighbors, and the number of their mutual butterflies will be ${c_d \choose 2}$, which makes $\beta(u)=\sum_{d \in D}{c_d \choose 2}$.
This phase has $O(\sum_{v \in V}{d(v)^2})$ time complexity, since a vertex $v \in V$ is accessed by each of its neighbors, and each time all the vertices in $N(v)$ are accessed.
In the worst case, it is $O(|U|^2)$, where the graph is a biclique.
We also used an ordering heuristic which is applied in the {\tt D2} and although it does not reduce the complexity, it enables to keep the $c$ and $D$ smaller, in line~\ref{ln:t3}, and results in less work in the following line. We omit the details for brevity.

\setlength{\textfloatsep}{0pt}
\begin{algorithm}[!t]
\small
\caption{\textsc{Wing Decomposition ($G=(U, V, E)$)}}
\label{alg:wingdecom}
\nonl \textcolor{black}{ \tcp{find the number of $\btrfly$ each edge $e \in E$ participates}}
$\beta(e) \leftarrow$ 0, butterfly counts $\forall~e \in E$\;\label{ln:w1}
\For{\meach $u \in U$ in order}{\label{ln:w2}
	\For{\meach $v_1, v_2~pair \in N(u)$}{\label{ln:w3}
		$I \leftarrow N(v_1) \cap N(v_2)$ s.t. $i \succ u~~\forall i \in I$\;\label{ln:w4}
		\For{\meach $i \in I$}{\label{ln:w5}
			$\beta(e)$++$,\forall e \in \btrfly(u, i, v_1, v_2)$\textcolor{black}{~\tcp*{each $\btrfly$ visited}}\label{ln:w6}
		}
	}
}
\nonl \textcolor{black}{\tcp{find the wing numbers of edges $e \in E$ by peeling}}
$\wing{e} \leftarrow$ wing numbers $\forall~e \in E$\;\label{ln:w7}
\For{\meach $e$~\textnormal{with minimum} $\beta(e)$~\textnormal{s.t.}~$\wing{e}$~\textnormal{is unassigned}}{\label{ln:w8}
  	$\wing{e} \leftarrow \beta(e)$\textcolor{black}{\tcp*{assign the wing number}}\label{ln:w9}
    	$\cB \leftarrow$ set  of $\btrfly$s containing $e$\;\label{ln:w10}
        \For{\meach $\btrfly \in \cB$}{\label{ln:w11}        
		\For{\meach~\textnormal{edge} $f$~\textnormal{in}~$\btrfly$~\textnormal{s.t.} $f \ne e$}{\label{ln:w12}
	    			\lIf{$\beta(f) > \beta(e)$}{$\beta(f)$- -\hfill\textcolor{black}{\tt ~// neigs updated}}\label{ln:w13}
		}
	}
	$E \leftarrow E \setminus e$\textcolor{black}{~\tcp*{edge is removed from the graph}}\label{ln:w14}
}
\Return array $\wing{\cdot}$\label{ln:w15}
\end{algorithm}

In the peeling process (lines~\ref{ln:t5} to~\ref{ln:t11}), we assign the tip numbers of the vertices, $\tip{\cdot}$, in a non-decreasing order.
We leverage a bucket data structure to efficiently retrieve the vertex with the fewest butterflies at each step.
Main operation at each step is to first assign the current butterfly number of $u$ as its tip number, find the vertices that has a common butterfly with $u$, and decrement their butterfly numbers, if larger than the last assigned tip number.
Since two vertices in $U$ can have multiple mutual butterflies, we find those counts and decrease them at once.
This again requires to collect the distance-2 neighbors in a multiset and we use {\tt D2} for that purpose in line~\ref{ln:t8}.
For each vertex that is not assigned a tip number yet, we decrease the butterfly number by $c \choose 2$, where $c$ is the cardinality of the common neighbors.
However, the butterfly number cannot be less than the last assigned tip number, and we check that in line~\ref{ln:t10}.
Time complexity is again characterized by $O(\sum_{v \in V}{d(v)^2})$ due to the {\tt D2} ($O(|U|^2)|$ in the worst case).
All of the additional data structures in both phases are in at most $O(|U|)$ size.

\vspace{-2ex}
\begin{theorem}
Given a bipartite graph $G=(U,V,E)$, Algorithm~\ref{alg:tipdecom} finds the tip numbers, $\tip{\cdot}$, of all $u \in U$ \textnormal{(Proof is omitted here, available in~\cite{longer}).}
\vspace{-1.5ex}
\end{theorem}

\begin{proof}
We define that $u, v \in U$ are friend vertices if they participate in a common butterfly.
$\tip{u}=k$ indicates that there are at least $k$ butterflies which contains $u$ and a friend vertex $v$ s.t. $\tip{v} \ge \tip{u}$.
This is enforced by the lines~\ref{ln:t10}-\ref{ln:t11}, where the butterfly count of a friend vertex is decreased if it will not be smaller than the tip number assigned at that step. In other words, any friend vertex with a smaller tip number does not contribute to the tip number of the vertex of interest. If Algorithm~\ref{alg:tipdecom} finds $\tip{u}=k$ for a vertex $u \in U$, then by Def.~\ref{def:tipno}, we need to show that (i) $\exists$ a $k$-tip $H \ni u$, (ii) $\nexists$ a $k^+$-tip $H \ni u$ ($k^+ > k$).\\
(i) Once $\tip{u}=k$ is found in Algorithm~\ref{alg:tipdecom}, stop and construct an induced subgraph $H \subset G$ by traversal as follows: Initially $H$ has only $u$. At each step, visit a vertex $v \in U$ s.t. $v$ participates in a butterfly with a vertex from $H$. If $\tip{v}$ is unassigned yet or equal to $k$, add $v$ to $H$. Continue the traversal until no such vertex $v$ can be found. At the end, $H$ is a $k$-tip since (1) each vertex has tip number of at least $k$, which means each has $\ge k$ butterflies, since the vertices are processed in the non-decreasing order of the butterfly counts (line~\ref{ln:t6}), (2) all the vertices are connected to each other with butterflies thanks to the butterfly based traversal operation, and (3) $H$ is maximal since it is the largest subgraph that can be found by the traversal.\\
(ii) $u$ cannot be in a $k^+$-tip. Assume it is. Then it should take part in at least $k^+$ butterflies and each butterfly contains a friend vertex that has at least $k^+$ butterflies. But, this implies that $\tip{u}=k^+$, contradiction.
\end{proof}

\vspace{-2ex}
\subsection{Wing decomposition: the Algorithm}

We apply a similar peeling approach to find the wing numbers of the edges in $E$, $\wing{\cdot}$.
Instead of looking at the vertex-butterfly relations, we investigate the involvements of edges in butterflies.
There are again two phases; counting the butterflies for each edge and the peeling process to find the wing numbers.
Locating the $k$-wings and building the hierarchy is again straightforward by the disjoint-set data structures~\cite{Sariyuce17} and not included here for brevity.

Algorithm~\ref{alg:wingdecom} outlines the \textsc{Wing Decomposition}.
Butterfly counting for each edge is done in lines~\ref{ln:w1} to~\ref{ln:w6}.
Note that it is different than the first phase of \textsc{Tip Decomposition}, and also more expensive, since we need to enumerate the butterflies to find the participating edges.
We compute intersections for each pair in the neighborhoods of vertices in $U$.
We utilize a total ordering of the vertices for efficient computation.
All the vertices $u \in U$ are processed in order (line~\ref{ln:w2}), and in each intersection operation, we only take the vertices that succeed $u$ (line~\ref{ln:w4}).
This enables to visit each butterfly only once.
Total complexity is $O(\sum_{u \in U}{\sum_{v_1, v_2 \in N(u)}{max ( d(v_1), d(v_2))}})$, and it is $O(|U||V||E|)$ in the worst case where the graph is a biclique.

In the peeling phase (lines~\ref{ln:w7} to~\ref{ln:w14}), wing numbers are assigned in a non-decreasing order.
Bucket structure helps to get the edge with the least number of butterflies at each step.
Similar to the \textsc{Tip Decomposition}, we assign the updated butterfly number of the edge $e$ as its wing number, say $k$, find the butterflies that $e$ participates, and decrement the butterfly counts of the other edges in those butterflies, if greater than $k$.
Note that there are four edges in a butterfly, thus we need to check the other three edges.
Lastly, we remove the edge $e$ from the graph.
Peeling phase has $O(\sum_{(u, v) \in E}{\sum_{w \in N(v)}{max (d(u), d(w))}})$ time complexity, and $O(|U||V||E|)$ in the worst case.
Additional space complexity is $O(|E|)$, since we store the butterfly and wing numbers for each edge.

\vspace{-1ex}
\begin{theorem}
Given a bipartite graph $G=(U,V,E)$, Algorithm~\ref{alg:wingdecom} finds the wing numbers, $\wing{\cdot}$, of all $e \in E$  \textnormal{(Proof is omitted here, available in~\cite{longer}).}
\end{theorem}
\begin{proof}
We define that $e, f \in E$ are friend edges if they participate in a common butterfly.
$\wing{e}=k$ indicates that there are at least $k$ butterflies which contains $e$ and three other friend edges $f,g,h$ s.t. $\wing{f} \ge \wing{e}$, $\wing{g} \ge \wing{e}$, and $\wing{h} \ge \wing{e}$.
This is enforced by the line~\ref{ln:w13}, where the butterfly count of an edge is decremented if larger than the wing number assigned at that step. In other words, any friend edge with a smaller wing number does not contribute to the wing number of the edge of interest.
If Algorithm~\ref{alg:wingdecom} finds $\wing{e}=k$ for a vertex $e \in E$, then by Def.~\ref{def:wingno}, we need to show that (i) $\exists$ a $k$-wing $H \ni e$, (ii) $\nexists$ a $k^+$-wing $H \ni e$ ($k^+ > k$).\\
(i) Once $\wing{e}=k$ is found in Algorithm~\ref{alg:wingdecom}, stop and take the remaining subgraph $H \subset G$. $H$ is a $k$-wing since (1) each edge has wing number of at least $k$, which means each has $\ge k$ butterflies, since the edges are processed in the non-decreasing order of their butterfly counts (line~\ref{ln:w8}), (2) all the edges are connected to each other with butterflies thanks to the butterfly based traversal operation, and (3) $H$ is maximal since it is the largest subgraph that can be found by the traversal.\\
(ii) $e$ cannot be in a $k^+$-wing. Assume it is. Then it should take part in at least $k^+$ butterflies and each butterfly contains neighbor edges with at least $k^+$ butterflies. But, this implies that $\wing{e}=k^+$, contradiction.
\end{proof}

\vspace{-3ex}
\section{Experiments}\label{sec:exps}

\noindent We evaluate our algorithms on real-world unweighted simple bipartite networks from SNAP~\cite{snap}, ICON~\cite{icon}, and Konect~\cite{konect}.
Table~\ref{tab:dataset} shows the important statistics for our dataset.
\con~is the author-paper network for the arXiv preprints about condensed matter physics, published between 1995 and 1999~\cite{Newman2001}.
\db~is another author-paper network that we constructed with the proceedings in the three top database conferences; VLDB, SIGMOD, and ICDE~\cite{dblp}.
\git~is the network between the users and repositories in the GitHub~\cite{github}.
\mar~is the occurrence relations between the Marvel characters and the comic books~\cite{marvel}.
\imdb~links the actors and the movies they played in~\cite{imdb}.
\dblp~is the author and paper network for all the papers in the DBLP website~\cite{dblp}.
\disaff~and \dissty~networks are obtained from an online music database~\cite{discog} and the former consists of the relations between artists and the production companies, whereas the latter is the network of the artists and the styles of their albums.
\itwiki~is the edit network of the Italian Wikipedia, having the users and the pages they edit.
\kindle~is the network between the Kindle books and the users who rated those books.
In Table~\ref{tab:dataset}, second to fourth columns show the number of primary vertices, secondary vertices, and edges in each network.
We assume that the primary vertices are the ones that drive the connections, which are authors, users, characters, actors, and artists.
In the fifth column, the number of edges in the projected graphs ($E_p$) are given.
We applied the projections as described in Def.~\ref{def:weproj}.
Last two columns are the butterfly ($\btrfly$) counts in each bipartite network and the triangle ($\triangle_p$) counts in the projected unipartite graph.
For the \disaff, \dissty, and \itwiki networks, projections cannot be computed in 36 hours, so we give lower bounds for their edge and triangle counts, based on the degrees of the secondary vertices.
Our implementation is in C++ and available~\footnote{\url{http://sariyuce.com/bnd.tar}}. We used gcc 5.2.0 at -O2 optimization level.
All experiments are performed on a Linux operating system running on a machine with Intel Xeon Haswell E5-2698 2.30 GHz processor with 128GB RAM.

\begin{table}[!t]
\centering
\small
\linespread{1}\selectfont
\renewcommand{\tabcolsep}{2pt}
\caption{\small \bf Statistics for the real-world bipartite graphs (in black) and their projections (in red). First three columns show the number of primary vertices, secondary vertices, and edges for the bipartite graphs, and the fifth column is the number of butterflies. Fourth and last columns (in red) are the number of edges and triangles in the projections, which applied on $U$.}
\vspace{-3ex}
\hspace{-2ex}
\begin{tabular}{|l||r|r|r|r|r|r|}\hline
\multicolumn{1}{|c||}{network} & \multicolumn{1}{c|}{$|U|$} & \multicolumn{1}{c|}{$|V|$} & \multicolumn{1}{c|}{$|E|$} &  \multicolumn{1}{c|}{\rd{$|E_p|$}} & \multicolumn{1}{c|}{$|\btrfly|$} & \multicolumn{1}{c|}{\rd{$|\triangle_{p}|$}} \\ \hline \hline
\con	&	$	16.73	$	~	K	&	$	22.02	$	~	K	&	$	58.60	$	~	K	&	\rd{	$	95.19	$	~	K	}	&	$	70.55	$	~	K	&	\rd{	$	68.04	$	~	K	}	\\	\hline
\db	&	$	11.19	$	~	K	&	$	8.92	$	~	K	&	$	30.72	$	~	K	&	\rd{	$	84.79	$	~	K	}	&	$	34.55	$	~	K	&	\rd{	$	95.66	$	~	K	}	\\	\hline
\git	&	$	56.56	$	~	K	&	$	123.35	$	~	K	&	$	440.24	$	~	K	&	\rd{	$	44.56	$	~	M	}	&	$	50.89	$	~	M	&	\rd{	$	962.55	$	~	M	}	\\	\hline
\mar	&	$	6.49	$	~	K	&	$	12.94	$	~	K	&	$	96.66	$	~	K	&	\rd{	$	336.53	$	~	K	}	&	$	10.71	$	~	M	&	\rd{	$	3.26	$	~	M	}	\\	\hline
\imdb	&	$	1.23	$	~	M	&	$	419.66	$	~	K	&	$	5.60	$	~	M	&	\rd{	$	157.56	$	~	M	}	&	$	42.49	$	~	M	&	\rd{	$	312.08	$	~	M	}	\\	\hline
\dblp	&	$	4.00	$	~	M	&	$	1.43	$	~	M	&	$	8.65	$	~	M	&	\rd{	$	315.89	$	~	M	}	&	$	21.04	$	~	M	&	\rd{	$	9.99	$	~	B	}	\\	\hline
\disaff	&	$	1.75	$	~	M	&	$	270.77	$	~	K	&	$	5.30	$	~	M	&	\rd{	$	> 7.20 	$	~	B	}	&	$	3.26	$	~	B	&	\rd{	$	> 167.99	$	~	T	}	\\	\hline
\dissty	&	$	1.62	$	~	M	&	$	0.38	$	~	K	&	$	5.74	$	~	M	&	\rd{	$	> 2.56 	$	~	T	}	&	$	77.38	$	~	B	&	\rd{	$	> 699.01 	$	~	Q	}	\\	\hline
\itwiki	&	$	2.26	$	~	M	&	$	137.69	$	~	K	&	$	12.64	$	~	M	&	\rd{	$	> 148.48 	$	~	B	}	&	$	298.49	$	~	B	&	\rd{	$	> 17.50 	$	~	Q	}	\\	\hline
\kindle	&	$	430.53	$	~	K	&	$	1.41	$	~	M	&	$	3.21	$	~	M	&	\rd{	$	93.19	$	~	M	}	&	$	15.51	$	~	M	&	\rd{	$	10.14	$	~	B	}	\\	\hline
\end{tabular}
\label{tab:dataset}
\end{table}

We compared the \textsc{TipDecomposition} (\textsc{Tip} in short, Algorithm~\ref{alg:tipdecom}) and the \textsc{WingDecomposition} (\textsc{Wing} in short, Algorithm~\ref{alg:wingdecom}) with the previous studies that find dense subgraphs with (or without) hierarchical relations in bipartite networks and or their projections.
\begin{compactitem}[\leftmargin=-1.5ex $\bullet$]
\item For the \textbf{unweighted projection}, we use two algorithms: $k$-core decomposition (Def.~\ref{def:kcore}) and $(2,3)$ nucleus decomposition (Def.~\ref{def:23}). $(2,3)$ nuclei subgraphs have been shown to be quite effective to find dense regions with detailed hierarchical relations~\cite{Huang14, Sariyuce15}.
\item For the \textbf{weighted projection}, we use the fractional $k$-cores~\cite{Giatsidis11}, described in Sec.~\ref{sec:rel}. To the best of our knowledge, it is the only peeling adaptation that works on weighted networks. It is designed to handle the bipartite author-paper networks by using their weighted projections.
\item Regarding the algorithms that directly focus on the \textbf{bipartite} data, Li \etal~\cite{Li13} proposed a $k$-truss adaptation, as explained at the end of Sec.~\ref{sec:rel}. Although the focus is on the bipartite connections, their algorithm relies on inserting edges between the vertices in the same set, and computes the $k$-trusses on those new edges and new triangles, which is essentially the same as the $(2,3)$ nucleus decomposition.
\item Apart from those, we also check the \textbf{$(P, Q)$-biclique densest subgraphs}, proposed by Mitzenmacher et al.~\cite{Babis15KDD}. We obtain the $(2,2)$-biclique (butterfly) densest region for each bipartite network in our dataset.
\end{compactitem}

For each bipartite subgraph, we report the size of primary and secondary vertex sets, and the edge density, i.e. $\frac{|E|}{|U|\cdot|V|}$.
For the $k$-core, fractional $k$-core, and  $(2,3)$ nucleus decompositions, we find the nuclei/cores in the projections ($G_p$), and then report the induced bipartite subgraphs using the vertices in those nuclei/cores.

\vspace{-2ex}
\subsection{Dense subgraph profiles}

\begin{figure*}[!b]
\centering
\captionsetup[subfigure]{captionskip=-1ex}
\vspace{-3.5ex}
\hspace{-5ex}
\subfloat[\small \textbf{\textsc{Wing}}]{\includegraphics[width=0.35\linewidth]{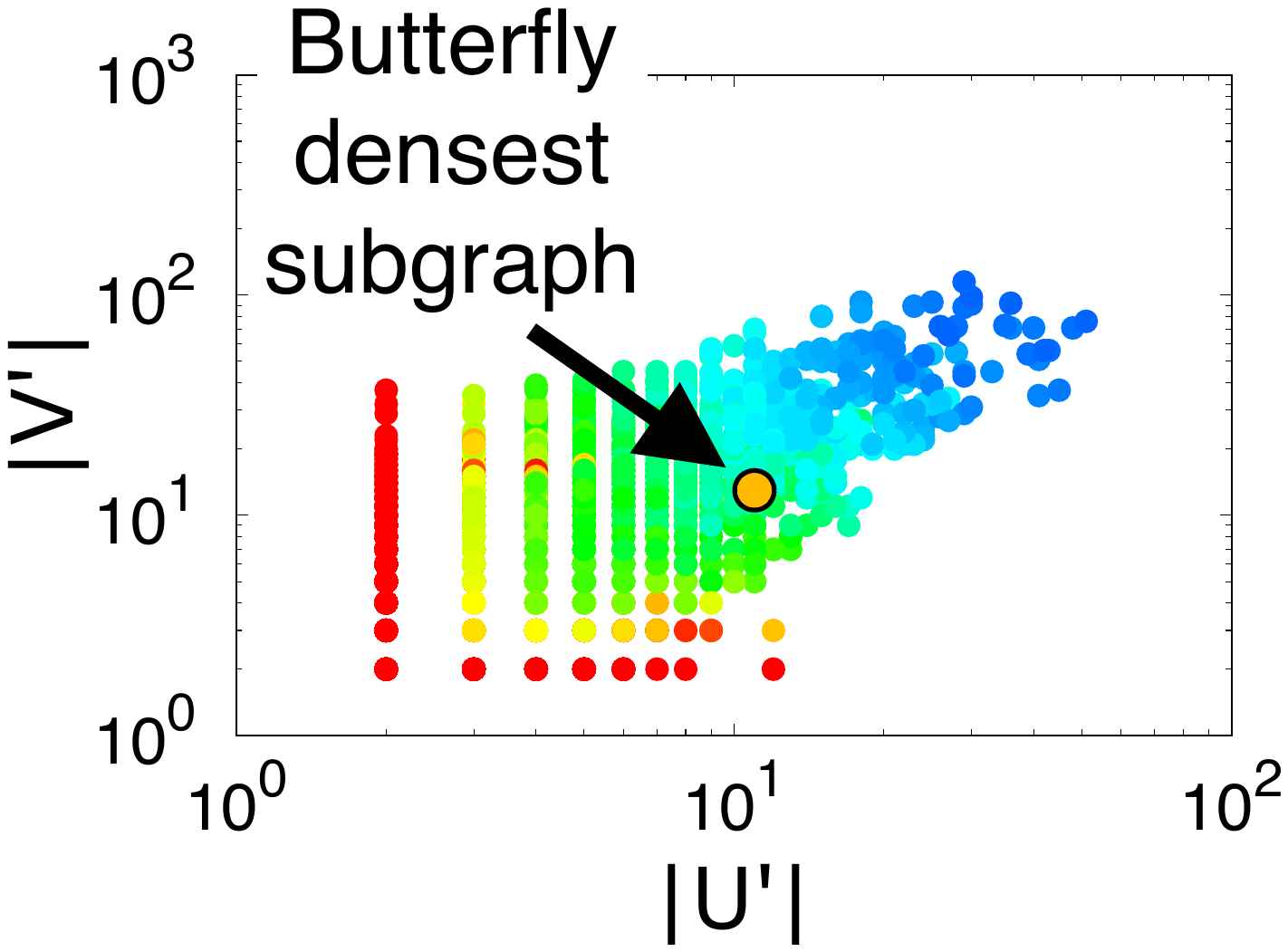}}
\hspace{-4.5ex}
\subfloat[\small \textbf{\textsc{$(2,3)$ nucleus}}]{\includegraphics[width=0.35\linewidth]{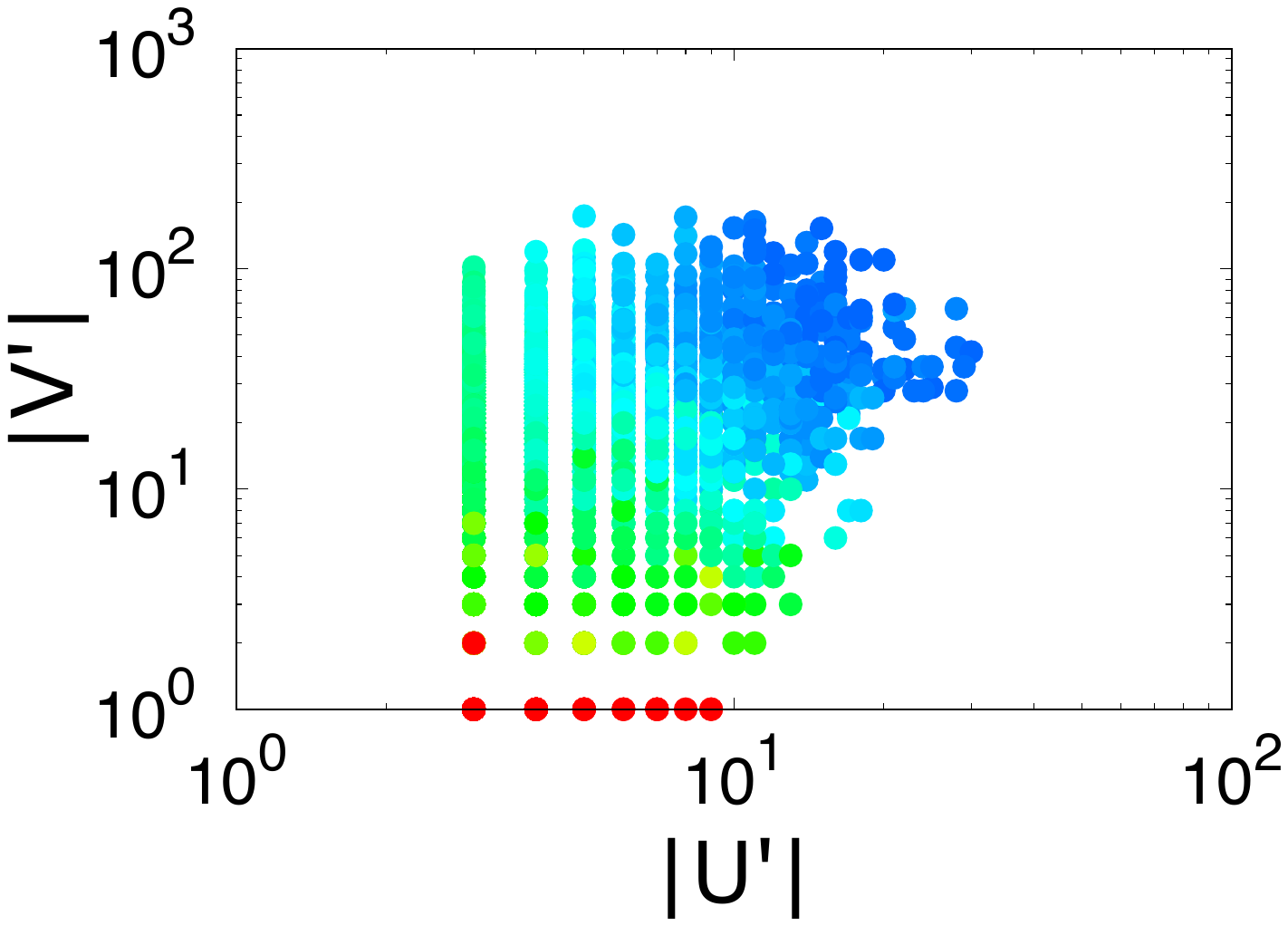}}
\hspace{-4.5ex}
\subfloat[\small \textbf{\textsc{Fractional $k$-core}}]{\includegraphics[width=0.35\linewidth]{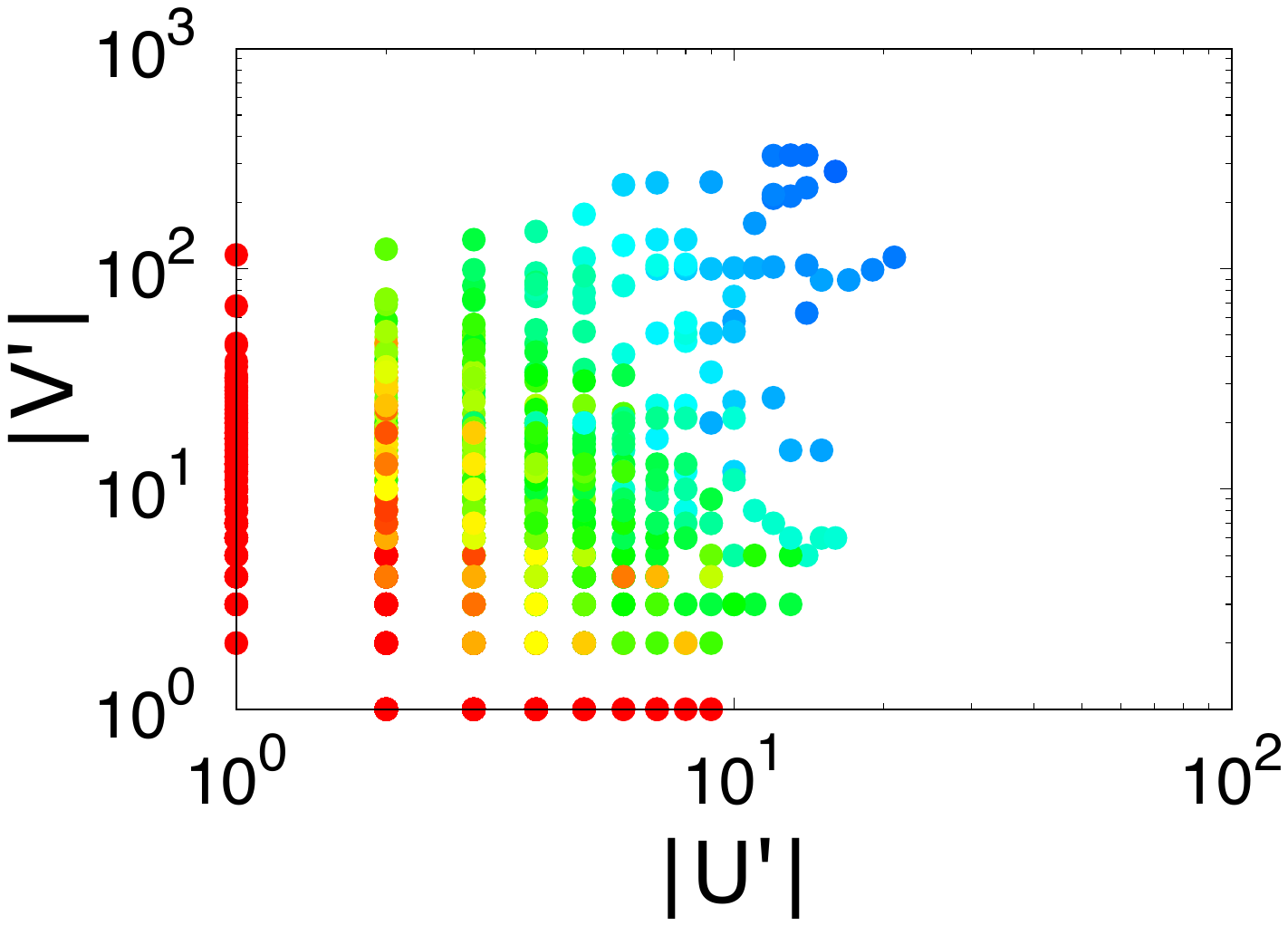}}
\hspace{-5ex}

\includegraphics[width=0.3\linewidth]{legend.pdf}
\hspace{-85ex}
\vspace{-1ex}
\caption{\small  \bf Dense subgraphs in \con. Each dot is a bipartite subgraph, the density, $|E'|/(|U'||V'|)$, is color coded, and $|U'|$ and $|V'|$ are given on the x- and y-axes. \textsc{Wing} results in 416 subgraphs with $\ge 0.5$ density and at least $5$ vertices in each side. Although it has many subgraphs with different qualities, none is as good as the butterfly densest subgraph, reported by~\protect{\cite{Babis15KDD}}. 
}
\label{fig:condmat}
\vspace{-3ex}
\end{figure*}

\begin{figure*}[!b]
\centering
\captionsetup[subfigure]{captionskip=-1ex}
\vspace{-4ex}
\hspace{-15ex}
\subfloat[\small \textbf{\textsc{Wing}}]{\includegraphics[width=0.35\linewidth]{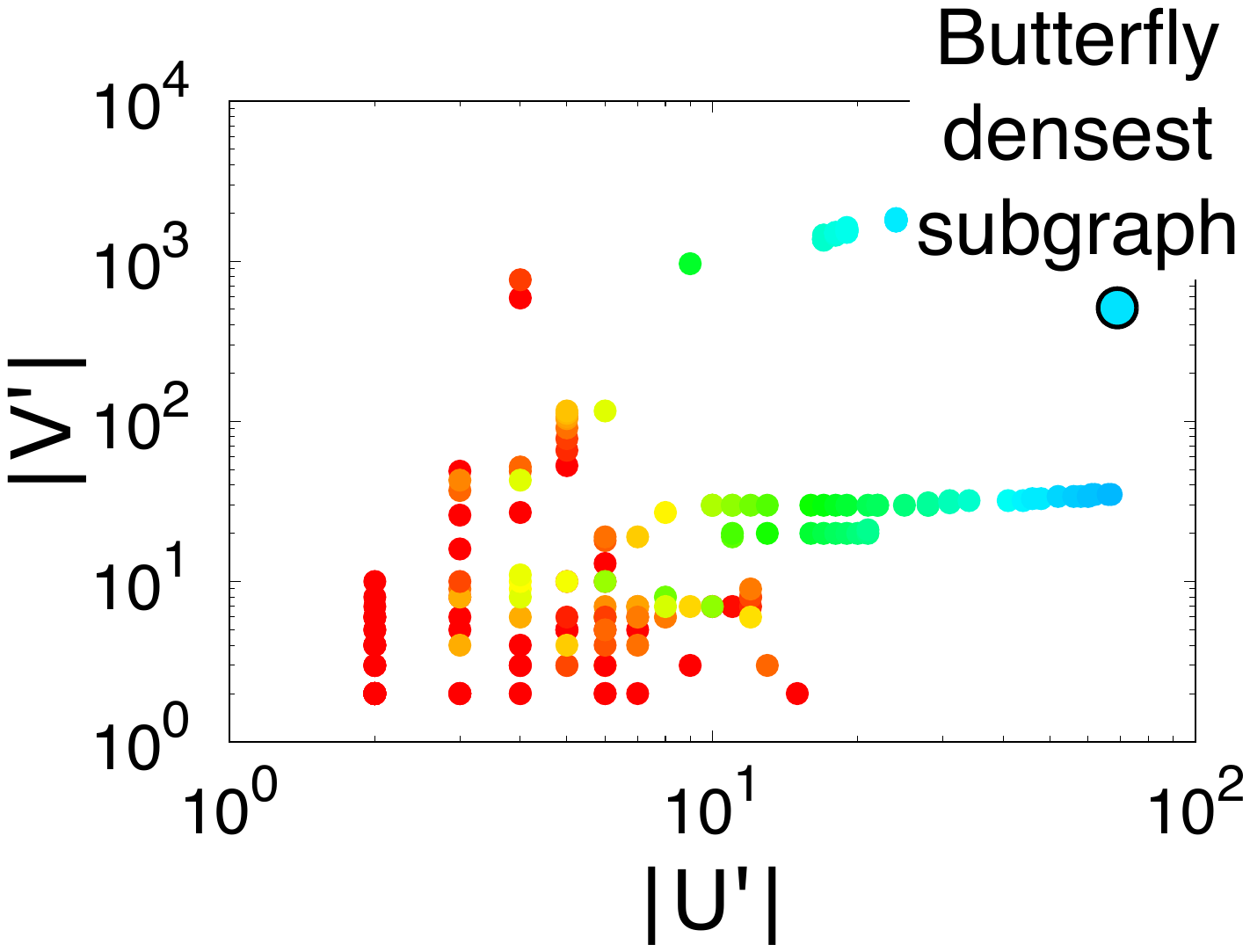}}
\hspace{-4.5ex}
\subfloat[\small \textbf{\textsc{Tip}}]{\includegraphics[width=0.35\linewidth]{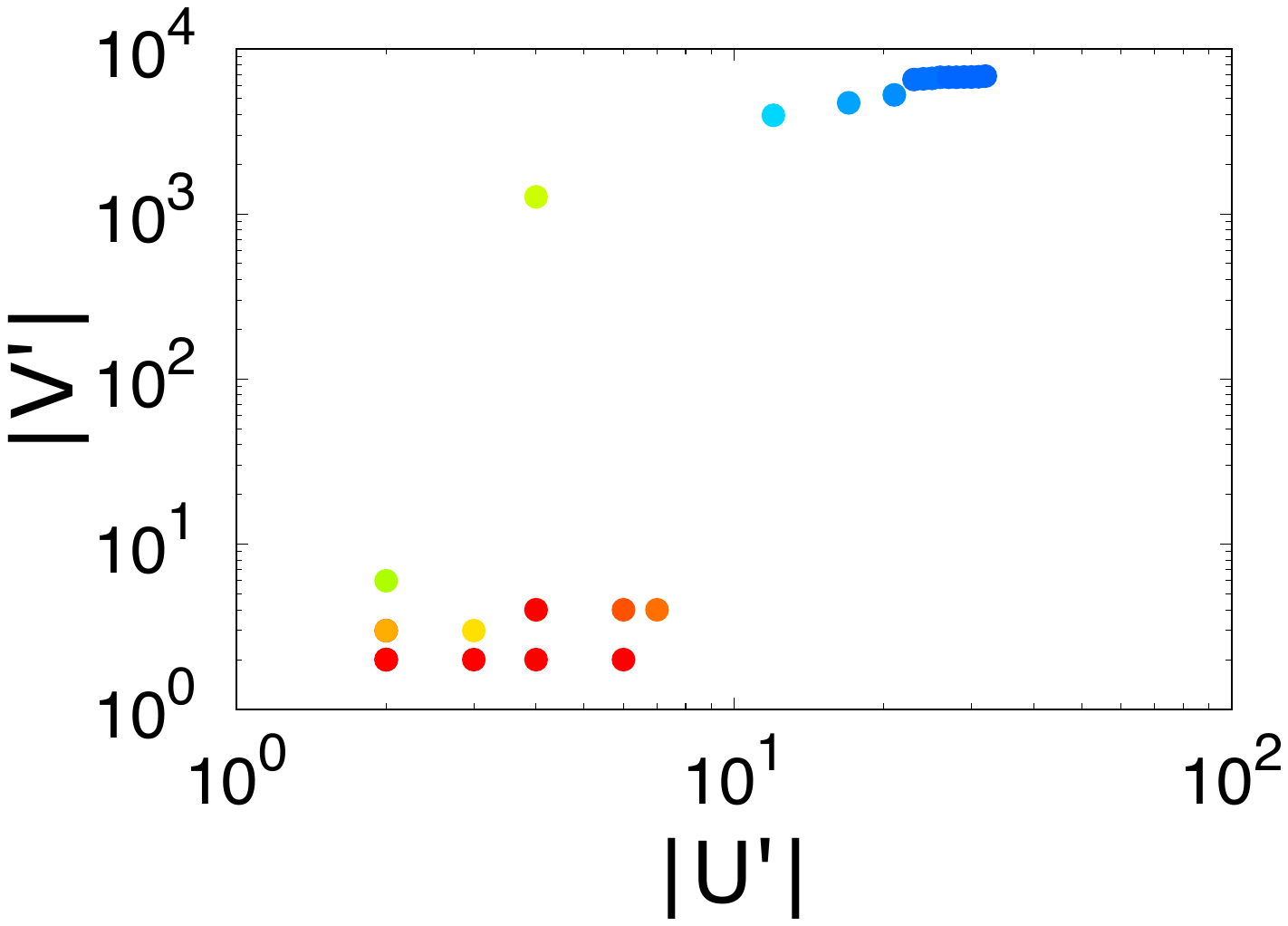}}
\hspace{-4.5ex}
\subfloat[\small \textbf{\textsc{$(2,3)$ nucleus}}]{\includegraphics[width=0.35\linewidth]{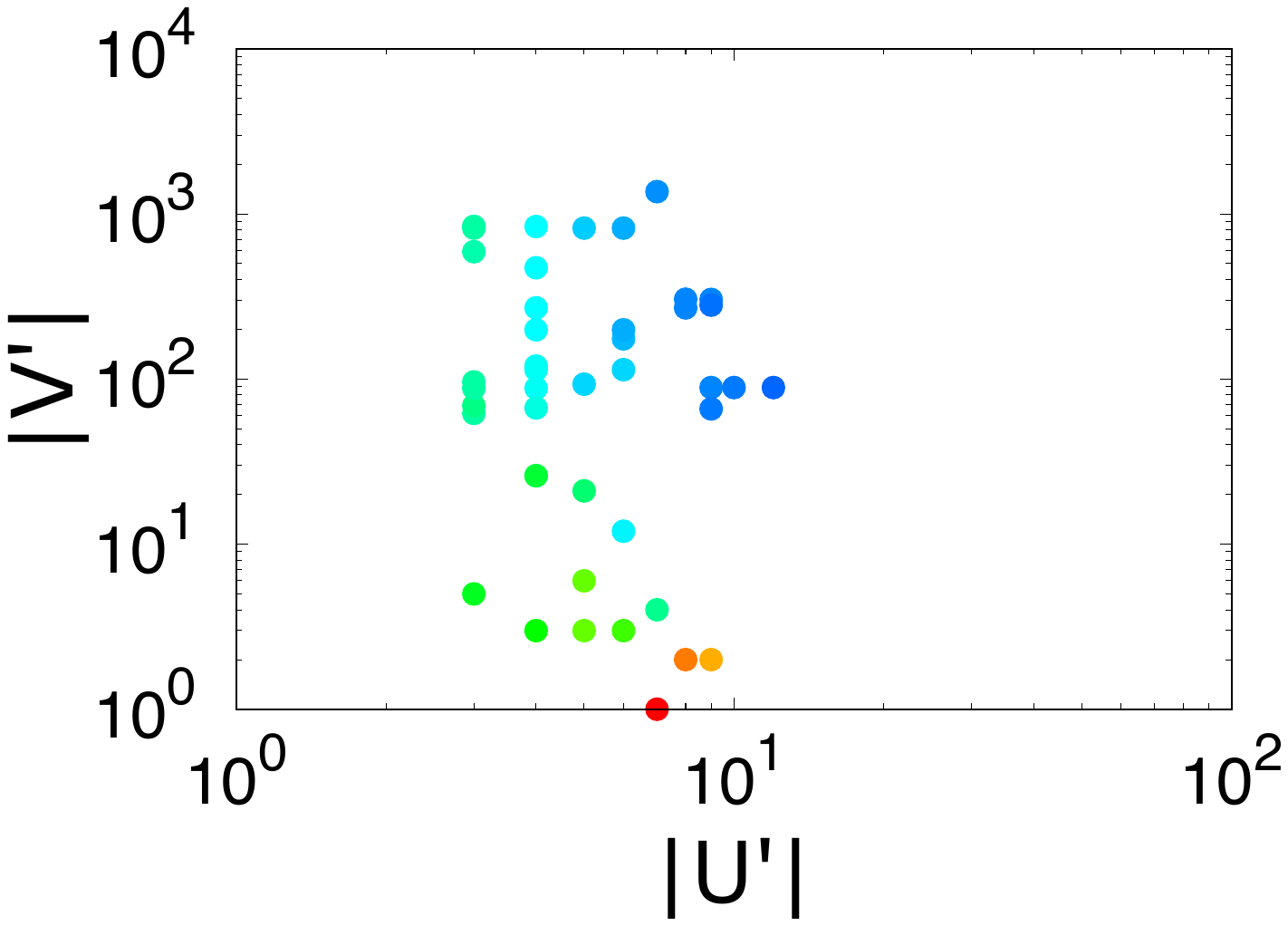}}
\hspace{-15ex}
\vspace{-3ex}
\caption{\small  \bf Dense subgraphs for \mar~network. \textsc{Wing} provides many subgraphs with $\ge 0.5$ density. $57$ of those have at least $5$ vertices in each side and $11$ have $10$ vertices in each. No other algorithm can get such subgraphs.}
\label{fig:marvel}
\end{figure*}

We compare the size and density of the bipartite subgraphs found by our algorithms and the previous works mentioned above.
Fig.~\ref{fig:condmat} (\con), ~\ref{fig:marvel} (\mar), 
and also~\ref{fig:introfig} (\imdb) (in Sec.~\ref{sec:intro}) summarize the results.
In all charts, each dot is a bipartite subgraph with at least $0.1$ edge density.
$|U|$ and $|V|$ are given on the x- and y-axes, and the density is color coded.

\begin{figure*}[!b]
\centering
\captionsetup[subfigure]{captionskip=-1.5ex}
\vspace{-4.35ex}
\hspace{-5ex}
\subfloat[\small \textbf{\textsc{Wing}}]{\includegraphics[width=0.35\linewidth]{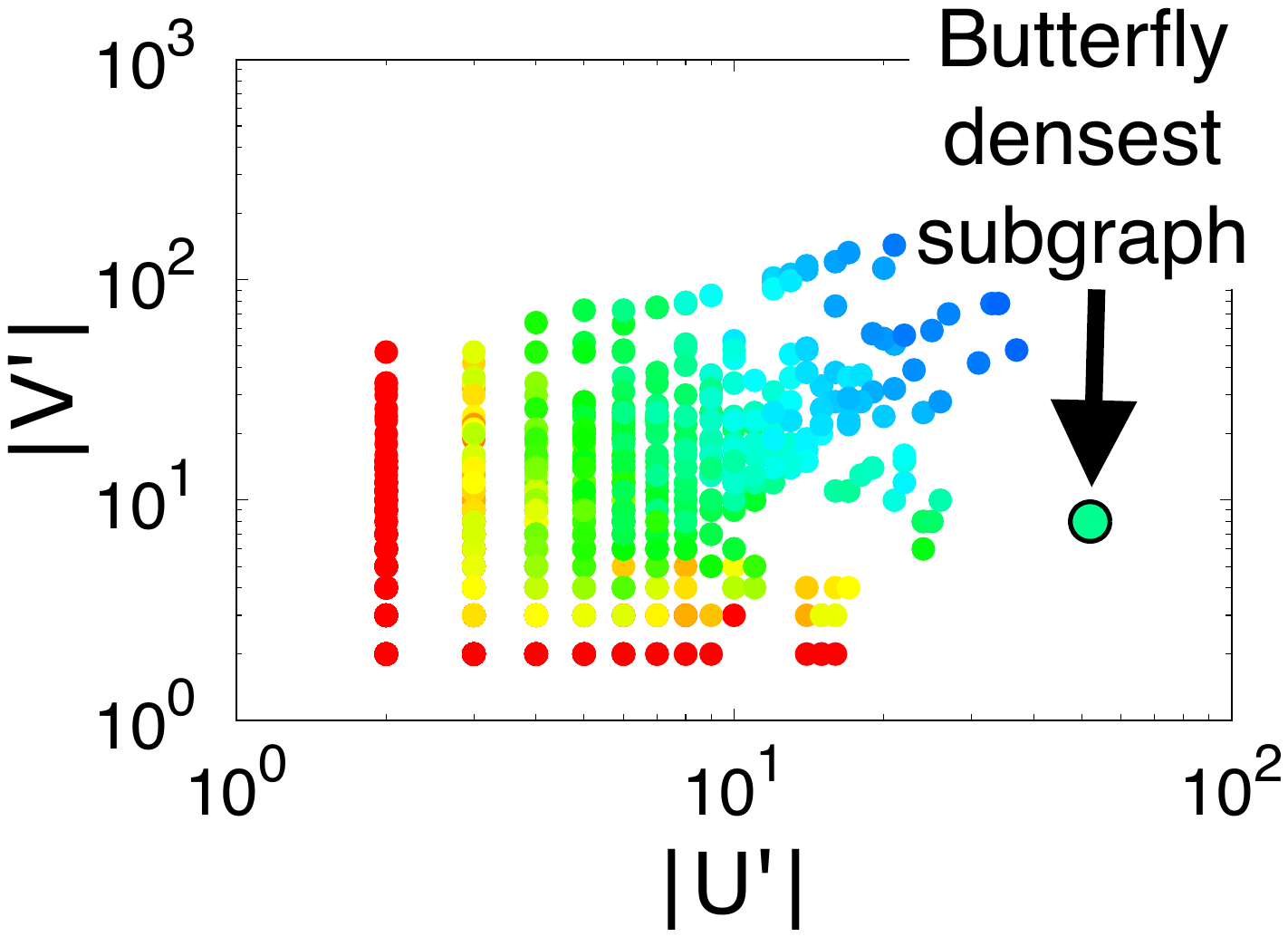}}
\hspace{-4.5ex}
\subfloat[\small \textbf{\textsc{Tip}}]{\includegraphics[width=0.35\linewidth]{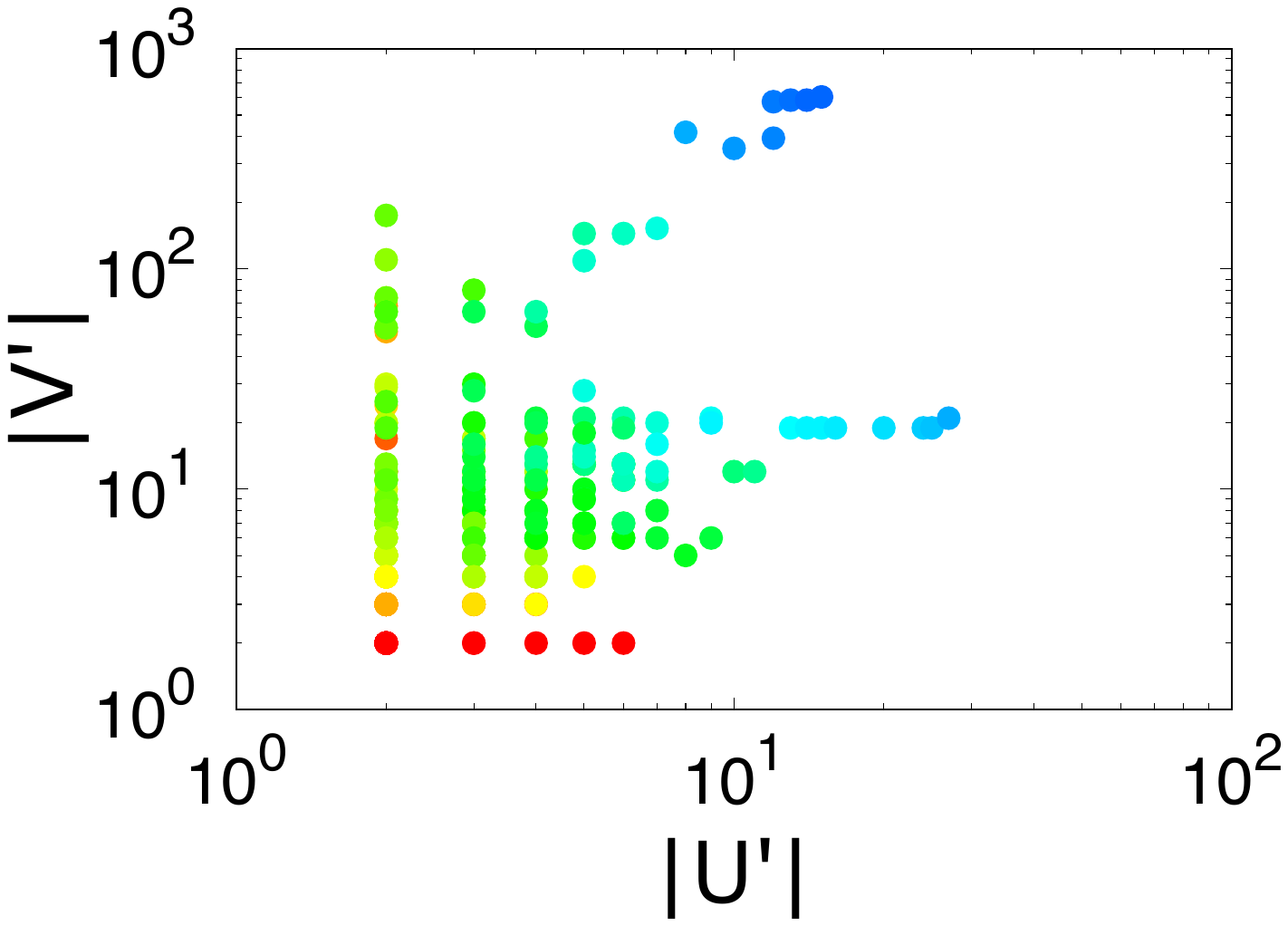}}
\hspace{-4.5ex}
\subfloat[\small \textbf{\textsc{$(2,3)$ nucleus}}]{\includegraphics[width=0.35\linewidth]{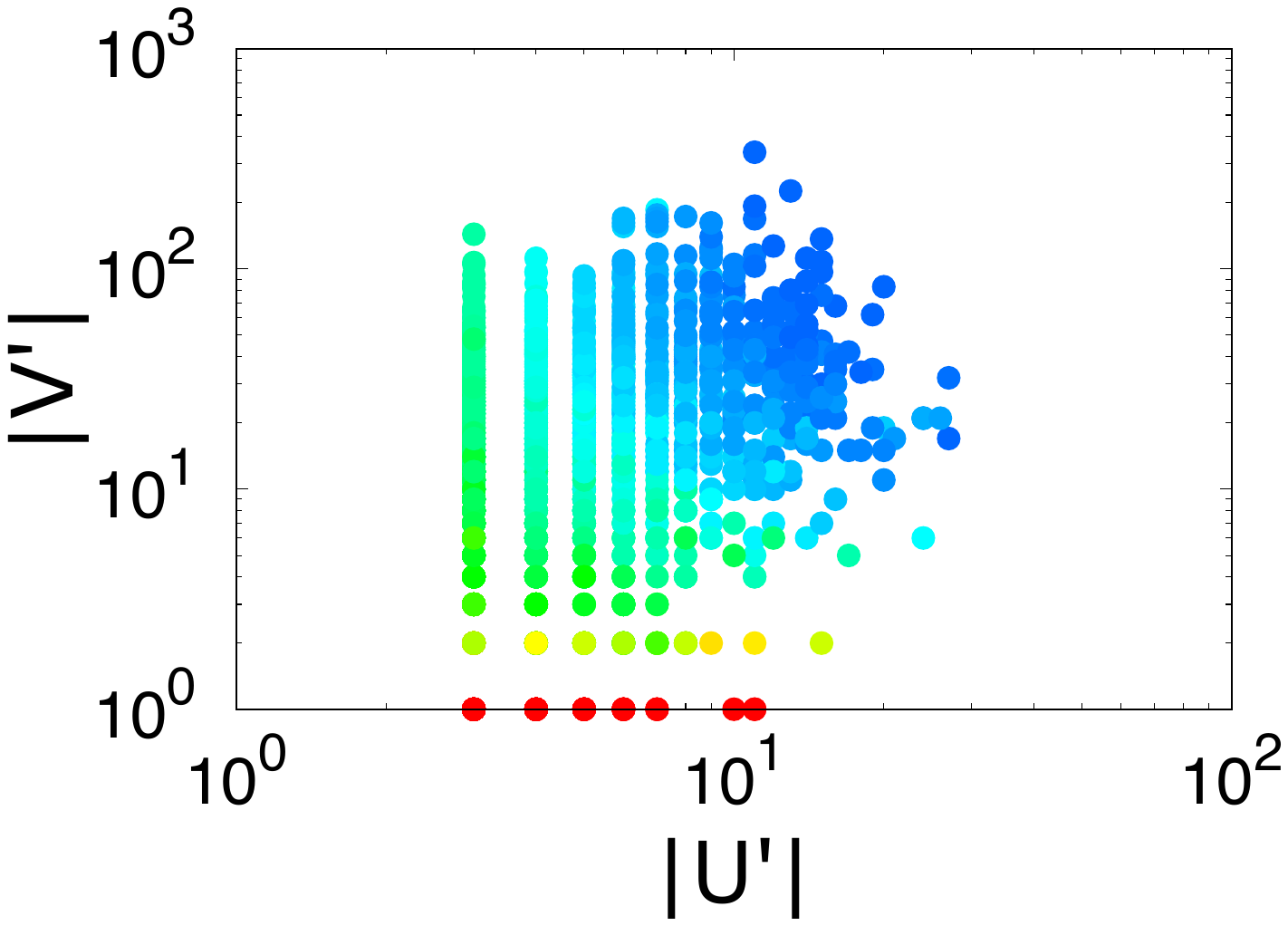}}
\hspace{-3ex}

\vspace{-3ex}
\caption{\small \db~network. Most dense structures in $(2,3)$ nuclei and fractional $k$-cores have only one vertex in either vertex set (the red dots along x- and y- axes).
Those subgraphs represent the collaborations of many authors in a single paper.
We observe that they are mostly papers on a software-product authored by a large group in a company.
In most cases those authors do not have any other papers, which makes the subgraph less informative.}
\label{fig:db}
\end{figure*}

Overall, we observe that many dense bipartite subgraphs with nontrivial sizes (on both sides) can be obtained with \textsc{Wing}. For \imdb~and \mar~networks, those subgraphs exhibit competitive quality (high density and large size) with respect to the butterfly densest subgraphs reported by~\cite{Babis15KDD}.
\textsc{Tip} also performs well on some instances with respect to other alternatives, but not as good as the \textsc{Wing}.
As we will show in Sec.~\ref{sec:runtime}, \textsc{Tip} is faster than the \textsc{Wing}, and the fair quality of the \textsc{Tip} can be preferred for applications with strict performance requirements.
We omit the results for $k$-cores since they consistently have lower densities and larger sizes than the \textsc{$(2,3)$ nucleus} and we observe that the densest subgraphs reported by the previous works are concentrated on the axes, meaning that they have a single vertex in either side, which is trivial.

For \con~network, in Fig.~\ref{fig:condmat}, \textsc{Wing} yields 416 subgraphs with at least 5 authors, 5 papers, and $0.5$ density. However, it cannot find a subgraph that is as good as the butterfly densest subgraph~\cite{Babis15KDD}, which has $11$ and $13$ vertices on each side with more than $0.8$ density.
\textsc{Tip} can find 59 such subgraphs, whereas the \textsc{$(2,3)$ nucleus} and \textsc{Fractional $k$-core} can only detect 14 and 20, respectively.
\textsc{Tip} and \textsc{Wing} are also effective to find detailed hierarchical relations; respectively, they report $77$ and $164$ subgraphs that contain multiple other subgraphs. $(2,3)$ nucleus can yield only $30$ such subgraphs and the other decompositions perform worse.


\mar~network, as seen in Fig.~\ref{fig:marvel}, shows more striking differences between \textsc{Wing} and the others; $42$ subgraphs appear with more than $0.7$ density and at least $5$ vertices on each side by using \textsc{Wing}, while no such subgraphs can be obtained with other algorithms, including \textsc{Tip}. Furthermore, \textsc{Wing} results in pretty competitive subgraphs with respect to the butterfly densest subgraph.

\subsection{Books in Amazon Kindle ratings}\label{sec:kindle}

We analyze the ratings data for the Amazon Kindle books.
The unweighted user-item bipartite graph has users on one side, items on the other side, and edges connect the users to the items they rated.
The first striking difference appears in the number of reported subgraphs: \textsc{Wing} gives $169$ distinct subgraphs that have at least $5$ vertices on each side and more than $0.5$ edge density. \textsc{Tip} reports $25$ such structures whereas \textsc{$(2,3)$ nucleus} only gives $6$.

\begin{figure*}[!t]
\vspace{-2ex}
\hspace{7ex}
\includegraphics[width=\linewidth]{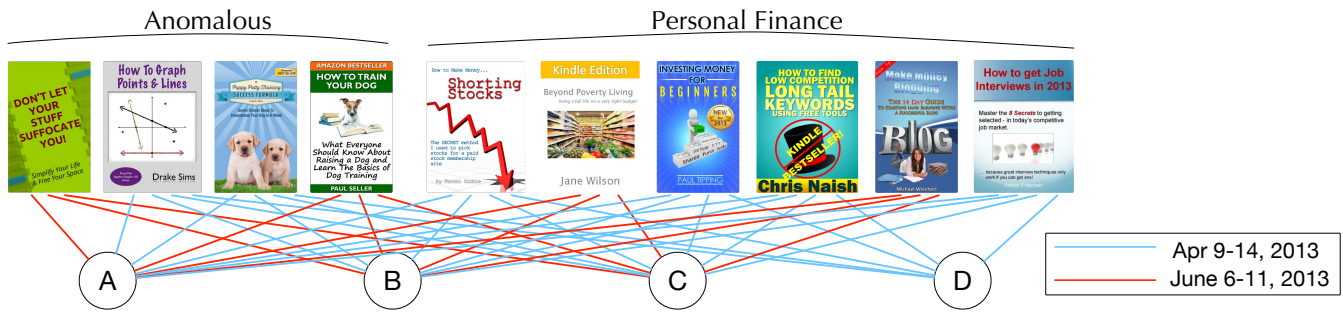}
\vspace{-2ex}
\caption{\small $10$ books rated by $4$ users, reported by \textsc{Wing}. $6$ books on the right are on personal finance whereas the $4$ on the left are unrelated. Further investigation reveals that those $4$ users rated the books in the same days and always gave $5$ stars, which are strong indicators for the fake reviews.}
\label{fig:10books}
\vspace{-6ex}
\hspace{-10ex}
\end{figure*}

\textbf{\textit{One example group that \textnormal{\textsc{Wing}} identifies while all the others cannot}} is a set of $38$ books that are mostly on the self-improvement theme ("how to" books, guides on relationships, healthy diets and personal finance) rated by $12$ users. 
Furthermore, there are three smaller groups within this group. The first has $6$ books and all are rated by $5$ users.
Those books are about the relationships between spouses, like ``\textit{Finding The Ins And Outs Of Relationships}''. 
Second group is a collection of $20$ books on guides for healthy diets and self-meditation. 
The third group, shown in Fig.~\ref{fig:10books}, reveals a different picture.
There are $10$ books that are rated by $4$ users, and $6$ of them are about the personal finance.
However, the other $4$ are totally unrelated; $2$ on dog training, $1$ on the elementary-level algebra and $1$ is about creating more space at home.
Checking the rating scores for those shows that that almost all the ratings are $5$ stars.
Furthermore, the dates those books are rated are very close: $2$ books are rated between April 9 and 14 in 2013, and the other $2$ are between June 6 and 11 in the same year.
Bursty reviews is an important indicator for the fake reviews, as \cite{Fei13} notes that ``\textit{Reviewers and reviews appearing in a burst are often related in the sense that spammers tend to work with other spammers and genuine reviewers tend to
appear together with other genuine reviewers}''.
Overall, \textsc{Wing} can reveal the sets of books on the same theme and also point to the anomalous behaviors, like the fake reviews, 
in bipartite networks. It does this by using only the graph topology \textit{without any metadata}.

\subsection{Authors in the top database conferences}\label{sec:database}

Here we highlight some interesting subgraphs and hierarchical structures found in the \db~network.
We report on what \textit{\textsc{Wing} and \textsc{Tip} can find but others cannot}, and also what \textit{we cannot find which can be identified by other algorithms}. 

Fig.~\ref{fig:db} shows the dense subgraph profiles for the \db~network.
Trends are similar to the other graphs explained above. \textsc{Wing} finds dense subgraphs with non-trivial sizes and also vying with the butterfly densest subgraph, \textsc{Tip} cannot perform as good as the \textsc{Wing}, and other algorithms mainly result in dense regions with a single vertex on either side, red dots along x- and y-axes.
In \textsc{$(2,3)$ nucleus} case, the red dots on the x-axis mostly correspond to the papers about a software product of a company and authored by a large group of researchers.
In most cases those authors do not have any other papers, thus no butterfly structures exist around them and cannot be found by our \textsc{Wing} and \textsc{Tip} algorithms.
One example is the paper entitled \textit{``Comdb2: Bloomberg's Highly Available Relational Database System''} in ICDE'10 is written by a large group of people in Bloomberg LP.
On the other hand, in the \textsc{Fractional $k$-core} case, there are also red dots appearing along the y-axis.
Those are the subgraphs with a single author and many papers.
However, they are isolated because the fractional $k$-core computation assigns a large weight to this vertex and there is no other vertex around with a close weight.
Divesh Srivastava\footnote{http://dblp.uni-trier.de/pers/hd/s/Srivastava:Divesh} is one such prolific researcher appearing on the y-axis with $144$ papers. But, there is no other dense subgraph in fractional $k$-cores that contains him, showing the weakness of the projection-based approach.
\textbf{\textit{Note that, none of those trivial subgraphs are reported by the \textnormal{\textsc{Tip}} or \textnormal{\textsc{Wing}}~algorithms, since no butterflies exist.}}
We also identified some structures that can be \textbf{\textit{only identified by the \textnormal{\textsc{Tip}} and/or \textnormal{\textsc{Wing}} algorithms}}.\\


\begin{figure*}[!t]
\centering
\hspace{-24ex}
\includegraphics[width=0.8\linewidth]{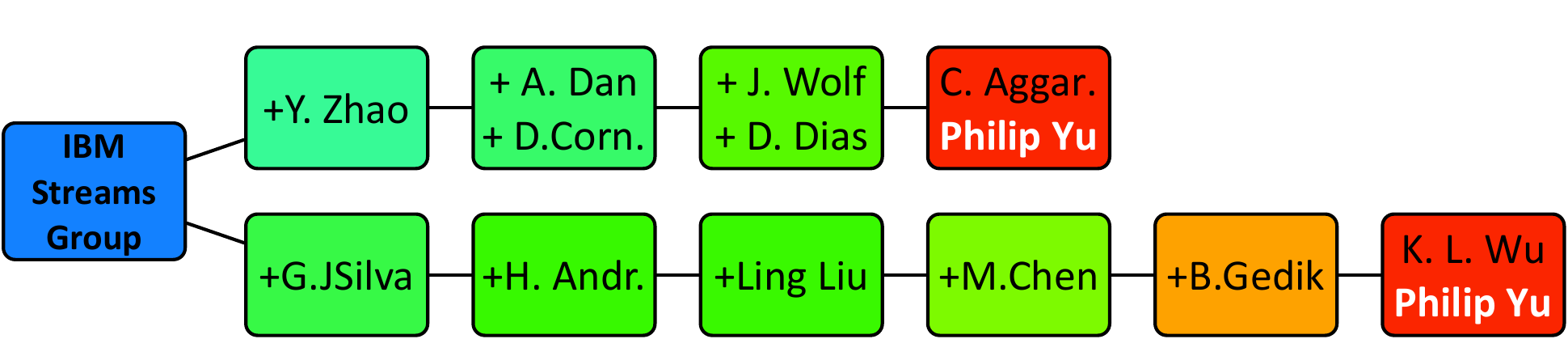}
\hspace{-25ex}
\vspace{-2ex}
\caption{\small The subgraphs reported by \textsc{Wing} in the \db~network that contains Philip Yu. Densities are color coded. Two chains show the collaborations in the IBM Research's Streams group, where the parent subgraph (left-most) contains $76$ researchers that contains staff members and intern students.}
\label{fig:phyu}
\end{figure*}

\vspace{-2ex}
\textbf{Philip Yu and IBM Research}: We also checked the densest subgraphs that contain a prolific researcher who is expected to participate in multiple structures.
We choose Philip Yu~\footnote{https://www.cs.uic.edu/PSYu/} for this purpose.
Tracing the hierarchical relations from his subgraph(s) in the \textsc{Tip} and \textsc{Wing} results gives interesting information about the Streams group at IBM Research, denoted in Fig.~\ref{fig:phyu}.
The bottom branch in the figure can be identified by both \textsc{Tip} and \textsc{Wing} whereas all other projection-based methods fail.
The densest subgraph on the right-most is a biclique, including Kun-Lung Wu, manager at IBM Research, and P. Yu, who used to work in the same group, and $13$ papers they co-authored.
They are contained in another subgraph which has $0.82$ density and includes Bu\u{g}ra Gedik (ex-staff member) in addition to K. L. Wu and P. Yu.
Looking at another parent subgraph in the chain, we find Ling Liu (B. Gedik's Ph.D. advisor) joins B. Gedik, K. L. Wu and P. Yu.
The further parent subgraphs in the chain reveals other researchers, H. Andrade and G. Jacques-Silva, who used to work in the same group.
\textsc{Wing} also reports another branch that contains P. Yu.
There is a biclique of Charu Aggarwal (staff member), P. Yu and their 14 papers.
Its parent subgraph has two more authors, Joel Wolf and Daniel Dias (staff members), and the further parent subgraphs in the chain include other people at the same institution.
This chain and the other chain described above merges in a subgraph that has 76 authors, who are the researchers and ex-interns in the Streams group.\footnote{All the information is obtained from the websites of the aforementioned people.}

\textbf{\textit{One weakness observed in the \textnormal{\textsc{Tip}} and \textnormal{\textsc{Wing}} is that they tend to identify pairs of authors first, and then enlarge those pairs to larger subgraphs.}} This results in long chains in the hierarchy which needs further filtering to extract large and dense structures. Main reason for this behavior is the quadratic increase in the number butterflies for author pairs with common papers, i.e., $n$ common neighbors gives $n\choose 2$ butterflies.

\vspace{-2ex}
\subsection{Runtime performances}\label{sec:runtime}

Lastly we check the runtimes.
Table~\ref{tab:runtimes} presents the results.
As pointed in Sec.~\ref{sec:alg}, \textsc{Wing} does more work than~\textsc{Tip} and it is also verified by the experiments; \textsc{Tip} can be up to $6$ times faster.
\textsc{$(2,3)$ nucleus} gives the most significant subgraphs among the projection based algorithms, however it has $O(|\triangle|)$ complexity \textit{on the projected graph} and suffers from the explosion in the number of triangles.
Note that, a $k$-clique is created for each secondary vertex with degree $k$, and the number of triangles increases \textbf{cubically},
for instance \dissty~network has less than $2$M vertices and around $5.5$M edges, but its projection has more than $699$ quadrillion triangles, which is impossible to process in a reasonable time.
Thus, \textsc{Wing} is orders of magnitude faster than the \textsc{$(2,3)$ nucleus}.
Regarding the densest butterfly subgraph~\cite{Babis15KDD} runtimes, we run~\cite{Babis15KDD} without sampling for all the graphs, except \disaff, \dissty, and \itwiki, and show the runtimes for butterfly enumeration and maximum flow computation phases.
Runtimes are mostly dominated by the maximum flow computation and significantly larger than the \textsc{Wing} runtimes.
For \disaff, \dissty, and \itwiki~graphs, the ones with the most butterflies, we set the sampling probabilities such that the maximum flow operation is run on $50$M butterflies (shown in red in Table~\ref{tab:runtimes}).
We observe that the butterfly enumeration, which does not depend on the sampling, takes the most time and it can be up to $33$\% of the total \textsc{Wing} time.
For the \itwiki~graph, it takes more than $24$K seconds to report a single densest butterfly subgraph while $97$K seconds is enough for the \textsc{Wing} to find many in a large spectrum.

\begin{table}[!t]
\centering
\vspace{1ex}
\small
\linespread{0}\selectfont
\linespread{1}\selectfont
\renewcommand{\tabcolsep}{2pt}
\caption{\small Runtimes of all the algorithms in seconds. \textsc{Wing} is orders of magnitude faster than the \textsc{$(2,3)$ nucleus} due to the increase in the number of triangles in the projection. \textsc{Core} and \textsc{Fractional core} are pretty fast, but not effective to report dense structures.}
\vspace{-2ex}
\begin{tabular}{|l|r|r|r|r|r||r|r|}\hline
& \multicolumn{2}{c|}{bipartite} & \multicolumn{3}{c||}{unipartite (projection)} & \multicolumn{2}{c|}{densest $\btrfly$~\cite{Babis15KDD}} \\
\multicolumn{1}{|c|}{(in sec.)} & \multicolumn{1}{c|}{\textsc{Tip}} & \multicolumn{1}{c|}{\textsc{Wing}} & \multicolumn{1}{c|}{\textsc{Core}} & \multicolumn{1}{c|}{\textsc{Fr. Core}} &  \multicolumn{1}{c||}{$(2,3)$ \textsc{nuc.}} &  \multicolumn{1}{c|}{enum} &  \multicolumn{1}{c|}{ max flow} \\ \hline \hline
%
\git	&	$	4.09	$	&	$	35.27	$	&	$	0.43	$	&	$	0.68	$	&	$	 121$,$643 	$	&	$	13.89	$	&	$	985	$	\\	\hline
\imdb	&	$	11.97	$	&	$	72.42	$	&	$	3.35	$	&	$	3.74	$	&	$	 8$,$831 	$	&	$	21.78	$	&	$	668	$	\\	\hline
\dblp	&	$	25.63	$	&	$	11.04	$	&	$	5.85	$	&	$	10.30	$	&	$	 9$,$378 	$	&	$	3.96	$	&	$	206	$	\\	\hline
\disaff	&	$	 1$,$328 	$	&	$	 3$,$789 	$	&	$	> 36 hrs	$	&	$	>36 hrs	$	&	$	 >36 hrs 	$	&	$	1$,$266	$	&	\rd{$1$,$146$}	\\	\hline
\dissty	&	$	 22$,$488 	$	&	$	 25$,$758 	$	&	$	> 36 hrs	$	&	$	>36 hrs	$	&	$	 >36 hrs 	$	&	$	7$,$264	$	&	\rd{$833$}	\\	\hline
\itwiki	&	$	 63$,$187 	$	&	$	 97$,$052 	$	&	$	> 36 hrs	$	&	$	>36 hrs	$	&	$	 >36 hrs 	$	&	$	23$,$564	$	&	\rd{$859$}	\\	\hline
\kindle	&	$	7.19	$	&	$	33.09	$	&	$	1.80	$	&	$	2.02	$	&	$	 29$,$057 	$	&	$	10.06	$	&	$	165	$	\\	\hline
\end{tabular}
\label{tab:runtimes}
\end{table}

\vspace{-2ex}
\section{Discussion}
\noindent Our algorithms for bipartite networks can find many dense substructures with the hierarchical relations.
Butterfly based definitions enable us to extract the meaningful regions in the graph and the use cases in Sec.~\ref{sec:database} and~\ref{sec:kindle} also verify this.
One weakness that we observed is that, number of butterflies per vertex or edge can quadratically increase when there is a large biclique. This results in dense but smaller subgraphs focusing around a few vertices.
Handling those regions is an interesting direction.
Another promising direction is to find efficient heuristics to count and enumerate the butterflies and other small bicliques.
There is a huge body of work on triangle counting and enumeration, but studies are immature for the bipartite structures.
Indeed, butterflies can easily reach to billions in a bipartite graph with a few million edges, whereas that many triangles in unipartite networks can be observed for hundreds of millions of edges.\\

\noindent\small{\textbf{Acknowledgements}: 
Sandia National Laboratories is a multimission laboratory managed and
operated by National Technology and Engineering Solutions of Sandia,
LLC., a wholly owned subsidiary of Honeywell International, Inc., for
the U.S. Department of Energy's National Nuclear Security Administration
under contract DE-NA-0003525.
This research used resources of the National Energy Research Scientific Computing Center, a DOE Office of Science User Facility supported by the Office of Science of the U.S. Department of Energy under Contract No. DE-AC02-05CH11231.\\
\vspace{-2ex}
}

\bibliographystyle{ACM-Reference-Format}
\bibliography{paper}


\begin{thebibliography}{00}


\ifx \showCODEN    \undefined \def \showCODEN     #1{\unskip}     \fi
\ifx \showDOI      \undefined \def \showDOI       #1{#1}\fi
\ifx \showISBNx    \undefined \def \showISBNx     #1{\unskip}     \fi
\ifx \showISBNxiii \undefined \def \showISBNxiii  #1{\unskip}     \fi
\ifx \showISSN     \undefined \def \showISSN      #1{\unskip}     \fi
\ifx \showLCCN     \undefined \def \showLCCN      #1{\unskip}     \fi
\ifx \shownote     \undefined \def \shownote      #1{#1}          \fi
\ifx \showarticletitle \undefined \def \showarticletitle #1{#1}   \fi
\ifx \showURL      \undefined \def \showURL       {\relax}        \fi
\providecommand\bibfield[2]{#2}
\providecommand\bibinfo[2]{#2}
\providecommand\natexlab[1]{#1}
\providecommand\showeprint[2][]{arXiv:#2}

\bibitem[\protect\citeauthoryear{??}{imd}{2016}]%
        {imdb}
 \bibinfo{year}{2016}\natexlab{}.
\newblock \bibinfo{title}{IMDb}.
\newblock   (\bibinfo{year}{2016}).
\newblock
\newblock
\shownote{({\tt www.imdb.com/interfaces}).}


\bibitem[\protect\citeauthoryear{??}{dis}{2017}]%
        {discog}
 \bibinfo{year}{2017}\natexlab{}.
\newblock \bibinfo{title}{Konect network dataset}.
\newblock   (\bibinfo{year}{2017}).
\newblock
\newblock
\shownote{({\tt http://www.discogs.com/}).}


\bibitem[\protect\citeauthoryear{Aksoy, Kolda, and Pinar}{Aksoy
  et~al\mbox{.}}{2017}]%
        {Aksoy16}
\bibfield{author}{\bibinfo{person}{S. Aksoy}, \bibinfo{person}{T.~G. Kolda},
  {and} \bibinfo{person}{A. Pinar}.} \bibinfo{year}{2017}\natexlab{}.
\newblock \showarticletitle{Measuring and Modeling Bipartite Graphs with
  Community Structure}.
\newblock \bibinfo{journal}{{\em Journal of Complex Networks\/}}
  \bibinfo{volume}{5}, \bibinfo{number}{4} (\bibinfo{year}{2017}),
  \bibinfo{pages}{581--603}.
\newblock


\bibitem[\protect\citeauthoryear{Alberich, Miro-Julia, and
  Rossell\'{o}}{Alberich et~al\mbox{.}}{2002}]%
        {marvel}
\bibfield{author}{\bibinfo{person}{R. Alberich}, \bibinfo{person}{J.
  Miro-Julia}, {and} \bibinfo{person}{F Rossell\'{o}}.}
  \bibinfo{year}{2002}\natexlab{}.
\newblock \showarticletitle{Marvel Universe looks almost like a real social
  network}.
\newblock \bibinfo{journal}{{\em CoRR\/}}
  \bibinfo{volume}{abs/cond-mat/0202174} (\bibinfo{year}{2002}).
\newblock


\bibitem[\protect\citeauthoryear{Batagelj and Zaversnik}{Batagelj and
  Zaversnik}{2002}]%
        {Batagelj02}
\bibfield{author}{\bibinfo{person}{Vladimir Batagelj} {and}
  \bibinfo{person}{Matjaz Zaversnik}.} \bibinfo{year}{2002}\natexlab{}.
\newblock \showarticletitle{Generalized Cores}.
\newblock \bibinfo{journal}{{\em CoRR\/}}  \bibinfo{volume}{cs.DS/0202039}
  (\bibinfo{year}{2002}).
\newblock


\bibitem[\protect\citeauthoryear{Batagelj and Zaversnik}{Batagelj and
  Zaversnik}{2003}]%
        {BaZa03}
\bibfield{author}{\bibinfo{person}{V. Batagelj} {and} \bibinfo{person}{M.
  Zaversnik}.} \bibinfo{year}{2003}\natexlab{}.
\newblock \showarticletitle{An O(m) Algorithm for Cores Decomposition of
  Networks}.
\newblock \bibinfo{journal}{{\em CoRR\/}}  \bibinfo{volume}{cs/0310049}
  (\bibinfo{year}{2003}).
\newblock


\bibitem[\protect\citeauthoryear{Benson, Gleich, and Leskovec}{Benson
  et~al\mbox{.}}{2016}]%
        {Benson16}
\bibfield{author}{\bibinfo{person}{A.~R. Benson}, \bibinfo{person}{D.~F.
  Gleich}, {and} \bibinfo{person}{J. Leskovec}.}
  \bibinfo{year}{2016}\natexlab{}.
\newblock \showarticletitle{Higher-order organization of complex networks}.
\newblock \bibinfo{journal}{{\em Science\/}} \bibinfo{volume}{353},
  \bibinfo{number}{6295} (\bibinfo{year}{2016}), \bibinfo{pages}{163--166}.
\newblock


\bibitem[\protect\citeauthoryear{Beutel, Xu, Guruswami, Palow, and
  Faloutsos}{Beutel et~al\mbox{.}}{2013}]%
        {Beutel13}
\bibfield{author}{\bibinfo{person}{A. Beutel}, \bibinfo{person}{W. Xu},
  \bibinfo{person}{V. Guruswami}, \bibinfo{person}{C. Palow}, {and}
  \bibinfo{person}{C. Faloutsos}.} \bibinfo{year}{2013}\natexlab{}.
\newblock \showarticletitle{CopyCatch: Stopping Group Attacks by Spotting
  Lockstep Behavior in Social Networks}. In \bibinfo{booktitle}{{\em
  Proceedings of the 22nd International Conference on World Wide Web}} {\em
  (\bibinfo{series}{WWW '13})}. \bibinfo{pages}{119--130}.
\newblock
\showISBNx{978-1-4503-2035-1}


\bibitem[\protect\citeauthoryear{Borgatti and Everett}{Borgatti and
  Everett}{1997}]%
        {Borgatti97}
\bibfield{author}{\bibinfo{person}{S.~P. Borgatti} {and} \bibinfo{person}{M.~G.
  Everett}.} \bibinfo{year}{1997}\natexlab{}.
\newblock \showarticletitle{Network analysis of 2-mode data}.
\newblock \bibinfo{journal}{{\em Social Networks\/}} \bibinfo{volume}{19},
  \bibinfo{number}{3} (\bibinfo{year}{1997}), \bibinfo{pages}{243 -- 269}.
\newblock


\bibitem[\protect\citeauthoryear{\c{C}ataly{\"{u}}rek and
  Aykanat}{\c{C}ataly{\"{u}}rek and Aykanat}{1999}]%
        {Umit99}
\bibfield{author}{\bibinfo{person}{{\"{U}}.~V. \c{C}ataly{\"{u}}rek} {and}
  \bibinfo{person}{C. Aykanat}.} \bibinfo{year}{1999}\natexlab{}.
\newblock \showarticletitle{Hypergraph-partitioning-based decomposition for
  parallel sparse-matrix vector multiplication}.
\newblock \bibinfo{journal}{{\em IEEE Transactions on Parallel and Distributed
  Systems\/}} \bibinfo{volume}{10}, \bibinfo{number}{7} (\bibinfo{year}{1999}),
  \bibinfo{pages}{673--693}.
\newblock


\bibitem[\protect\citeauthoryear{Cerinsek and Batagelj}{Cerinsek and
  Batagelj}{2015}]%
        {Cerinsek15}
\bibfield{author}{\bibinfo{person}{M. Cerinsek} {and} \bibinfo{person}{V.
  Batagelj}.} \bibinfo{year}{2015}\natexlab{}.
\newblock \showarticletitle{Generalized two-mode cores}.
\newblock \bibinfo{journal}{{\em Social Networks\/}}  \bibinfo{volume}{42}
  (\bibinfo{year}{2015}), \bibinfo{pages}{80 -- 87}.
\newblock


\bibitem[\protect\citeauthoryear{Chacon}{Chacon}{2009}]%
        {github}
\bibfield{author}{\bibinfo{person}{S. Chacon}.}
  \bibinfo{year}{2009}\natexlab{}.
\newblock \bibinfo{title}{The 2009 GitHub Contest}.
\newblock   (\bibinfo{year}{2009}).
\newblock
\newblock
\shownote{({\tt github.com/blog/466-the-2009-github-contest}).}


\bibitem[\protect\citeauthoryear{Chen and Saad}{Chen and Saad}{2012}]%
        {Chen12}
\bibfield{author}{\bibinfo{person}{J. Chen} {and} \bibinfo{person}{Y. Saad}.}
  \bibinfo{year}{2012}\natexlab{}.
\newblock \showarticletitle{Dense Subgraph Extraction with Application to
  Community Detection}.
\newblock \bibinfo{journal}{{\em IEEE Transactions on Knowledge \& Data
  Engineering\/}} \bibinfo{volume}{24}, \bibinfo{number}{7}
  (\bibinfo{year}{2012}), \bibinfo{pages}{1216--1230}.
\newblock
\showISSN{1041-4347}


\bibitem[\protect\citeauthoryear{Clauset, Tucker, and Sainz}{Clauset
  et~al\mbox{.}}{2016}]%
        {icon}
\bibfield{author}{\bibinfo{person}{A. Clauset}, \bibinfo{person}{E. Tucker},
  {and} \bibinfo{person}{M. Sainz}.} \bibinfo{year}{2016}\natexlab{}.
\newblock \bibinfo{title}{The Colorado Index of Complex Networks.}
\newblock   (\bibinfo{year}{2016}).
\newblock
\newblock
\shownote{({\tt icon.colorado.edu}).}


\bibitem[\protect\citeauthoryear{Cohen}{Cohen}{2008}]%
        {Cohen08}
\bibfield{author}{\bibinfo{person}{J. Cohen}.} \bibinfo{year}{2008}\natexlab{}.
\newblock \showarticletitle{Trusses: Cohesive subgraphs for social network
  analysis}.
\newblock   \bibinfo{volume}{National Security Agency Technical Report}
  (\bibinfo{year}{2008}).
\newblock


\bibitem[\protect\citeauthoryear{Dhillon}{Dhillon}{2001}]%
        {Dhillon01}
\bibfield{author}{\bibinfo{person}{I.S. Dhillon}.}
  \bibinfo{year}{2001}\natexlab{}.
\newblock \showarticletitle{Co-clustering Documents and Words Using Bipartite
  Spectral Graph Partitioning}. In \bibinfo{booktitle}{{\em Proceedings of the
  Seventh ACM SIGKDD International Conference on Knowledge Discovery and Data
  Mining}} {\em (\bibinfo{series}{KDD '01})}. \bibinfo{pages}{269--274}.
\newblock
\showISBNx{1-58113-391-X}


\bibitem[\protect\citeauthoryear{Everett and Borgatti}{Everett and
  Borgatti}{2013}]%
        {Everett13}
\bibfield{author}{\bibinfo{person}{M.G. Everett} {and} \bibinfo{person}{S.P.
  Borgatti}.} \bibinfo{year}{2013}\natexlab{}.
\newblock \showarticletitle{The dual-projection approach for two-mode
  networks}.
\newblock \bibinfo{journal}{{\em Social Networks\/}} \bibinfo{volume}{35},
  \bibinfo{number}{2} (\bibinfo{year}{2013}), \bibinfo{pages}{204 -- 210}.
\newblock


\bibitem[\protect\citeauthoryear{Fain and Pedersen}{Fain and Pedersen}{2006}]%
        {Fain06}
\bibfield{author}{\bibinfo{person}{D.~C. Fain} {and} \bibinfo{person}{J.~O.
  Pedersen}.} \bibinfo{year}{2006}\natexlab{}.
\newblock \showarticletitle{Sponsored search: A brief history}.
\newblock \bibinfo{journal}{{\em Bulletin of the American Society for
  Information Science and Technology\/}} \bibinfo{volume}{32},
  \bibinfo{number}{2} (\bibinfo{year}{2006}), \bibinfo{pages}{12--13}.
\newblock
\showISSN{1550-8366}


\bibitem[\protect\citeauthoryear{Fei, Mukherjee, Liu, Hsu, Castellanos, and
  Ghosh}{Fei et~al\mbox{.}}{2013}]%
        {Fei13}
\bibfield{author}{\bibinfo{person}{G. Fei}, \bibinfo{person}{A. Mukherjee},
  \bibinfo{person}{B. Liu}, \bibinfo{person}{M. Hsu}, \bibinfo{person}{M.
  Castellanos}, {and} \bibinfo{person}{R Ghosh}.}
  \bibinfo{year}{2013}\natexlab{}.
\newblock \showarticletitle{Exploiting Burstiness in Reviews for Review Spammer
  Detection}. In \bibinfo{booktitle}{{\em ICWSM}}. \bibinfo{publisher}{The AAAI
  Press}.
\newblock
\showISBNx{978-1-57735-610-3}


\bibitem[\protect\citeauthoryear{Giatsidis, Thilikos, and
  Vazirgiannis}{Giatsidis et~al\mbox{.}}{2011}]%
        {Giatsidis11}
\bibfield{author}{\bibinfo{person}{C. Giatsidis}, \bibinfo{person}{D.~M.
  Thilikos}, {and} \bibinfo{person}{M. Vazirgiannis}.}
  \bibinfo{year}{2011}\natexlab{}.
\newblock \showarticletitle{Evaluating Cooperation in Communities with the
  $k$-Core Structure}. In \bibinfo{booktitle}{{\em ASONAM}}.
  \bibinfo{pages}{87--93}.
\newblock


\bibitem[\protect\citeauthoryear{Gibson, Kumar, and Tomkins}{Gibson
  et~al\mbox{.}}{2005}]%
        {Gibson05}
\bibfield{author}{\bibinfo{person}{D. Gibson}, \bibinfo{person}{R. Kumar},
  {and} \bibinfo{person}{A. Tomkins}.} \bibinfo{year}{2005}\natexlab{}.
\newblock \showarticletitle{Discovering Large Dense Subgraphs in Massive
  Graphs}. In \bibinfo{booktitle}{{\em VLDB}}. \bibinfo{pages}{721--732}.
\newblock


\bibitem[\protect\citeauthoryear{Gregori, Lenzini, and Orsini}{Gregori
  et~al\mbox{.}}{2011}]%
        {Gregori11}
\bibfield{author}{\bibinfo{person}{E. Gregori}, \bibinfo{person}{L. Lenzini},
  {and} \bibinfo{person}{C. Orsini}.} \bibinfo{year}{2011}\natexlab{}.
\newblock \showarticletitle{k-dense communities in the internet AS-level
  topology}. In \bibinfo{booktitle}{{\em International Conf. on Communication
  Systems and Networks (COMSNETS)}}. \bibinfo{pages}{1--10}.
\newblock


\bibitem[\protect\citeauthoryear{Huang, Cheng, Qin, Tian, and Yu}{Huang
  et~al\mbox{.}}{2014}]%
        {Huang14}
\bibfield{author}{\bibinfo{person}{X. Huang}, \bibinfo{person}{H. Cheng},
  \bibinfo{person}{L. Qin}, \bibinfo{person}{W. Tian}, {and}
  \bibinfo{person}{J.~X. Yu}.} \bibinfo{year}{2014}\natexlab{}.
\newblock \showarticletitle{Querying K-truss Community in Large and Dynamic
  Graphs}. In \bibinfo{booktitle}{{\em Proc. of the ACM SIGMOD International
  Conf. on Management of Data}}. \bibinfo{pages}{1311--1322}.
\newblock


\bibitem[\protect\citeauthoryear{Kumar, Raghavan, Rajagopalan, and
  Tomkins}{Kumar et~al\mbox{.}}{1999}]%
        {Kumar99}
\bibfield{author}{\bibinfo{person}{R. Kumar}, \bibinfo{person}{P. Raghavan},
  \bibinfo{person}{S. Rajagopalan}, {and} \bibinfo{person}{A. Tomkins}.}
  \bibinfo{year}{1999}\natexlab{}.
\newblock \showarticletitle{Trawling the Web for Emerging Cyber-communities}.
  In \bibinfo{booktitle}{{\em WWW}}. \bibinfo{pages}{1481--1493}.
\newblock


\bibitem[\protect\citeauthoryear{Kunegis}{Kunegis}{2017}]%
        {konect}
\bibfield{author}{\bibinfo{person}{Jerome Kunegis}.}
  \bibinfo{year}{2017}\natexlab{}.
\newblock \bibinfo{title}{Konect network dataset}.
\newblock   (\bibinfo{year}{2017}).
\newblock
\newblock
\shownote{({\tt http://konect.uni-koblenz.de}).}


\bibitem[\protect\citeauthoryear{Latapy, Magnien, and Vecchio}{Latapy
  et~al\mbox{.}}{2008}]%
        {Latapy08}
\bibfield{author}{\bibinfo{person}{M. Latapy}, \bibinfo{person}{C. Magnien},
  {and} \bibinfo{person}{N.~Del Vecchio}.} \bibinfo{year}{2008}\natexlab{}.
\newblock \showarticletitle{Basic notions for the analysis of large two-mode
  networks}.
\newblock \bibinfo{journal}{{\em Social Networks\/}} \bibinfo{volume}{30},
  \bibinfo{number}{1} (\bibinfo{year}{2008}), \bibinfo{pages}{31 -- 48}.
\newblock


\bibitem[\protect\citeauthoryear{Leskovec and Krevl}{Leskovec and
  Krevl}{2014}]%
        {snap}
\bibfield{author}{\bibinfo{person}{Jure Leskovec} {and} \bibinfo{person}{Andrej
  Krevl}.} \bibinfo{year}{2014}\natexlab{}.
\newblock \bibinfo{title}{{SNAP Datasets}: {Stanford} Large Network Dataset
  Collection}.
\newblock   (\bibinfo{date}{June} \bibinfo{year}{2014}).
\newblock
\newblock
\shownote{({\tt snap.stanford.edu/data}).}


\bibitem[\protect\citeauthoryear{Ley}{Ley}{2016}]%
        {dblp}
\bibfield{author}{\bibinfo{person}{Michael Ley}.}
  \bibinfo{year}{2016}\natexlab{}.
\newblock \bibinfo{title}{DBLP computer science bibliography}.
\newblock   (\bibinfo{date}{Sept.} \bibinfo{year}{2016}).
\newblock
\newblock
\shownote{({\tt dblp.uni-trier.de}).}


\bibitem[\protect\citeauthoryear{Li, Kuboyama, and Sakamoto}{Li
  et~al\mbox{.}}{2013}]%
        {Li13}
\bibfield{author}{\bibinfo{person}{Y. Li}, \bibinfo{person}{T. Kuboyama}, {and}
  \bibinfo{person}{H. Sakamoto}.} \bibinfo{year}{2013}\natexlab{}.
\newblock \showarticletitle{Truss Decomposition for Extracting Communities in
  Bipartite Graph}. In \bibinfo{booktitle}{{\em IMMM 2013 : The Third
  International Conference on Advances in Information Mining and Management}}.
\newblock


\bibitem[\protect\citeauthoryear{Matula and Beck}{Matula and Beck}{1983}]%
        {MaBe83}
\bibfield{author}{\bibinfo{person}{D. Matula} {and} \bibinfo{person}{L. Beck}.}
  \bibinfo{year}{1983}\natexlab{}.
\newblock \showarticletitle{Smallest-last ordering and clustering and graph
  coloring algorithms}.
\newblock \bibinfo{journal}{{\em Journal of ACM\/}} \bibinfo{volume}{30},
  \bibinfo{number}{3} (\bibinfo{year}{1983}), \bibinfo{pages}{417--427}.
\newblock


\bibitem[\protect\citeauthoryear{Mitzenmacher, Pachocki, Peng, Tsourakakis, and
  Xu}{Mitzenmacher et~al\mbox{.}}{2015}]%
        {Babis15KDD}
\bibfield{author}{\bibinfo{person}{M. Mitzenmacher}, \bibinfo{person}{J.
  Pachocki}, \bibinfo{person}{R. Peng}, \bibinfo{person}{C. Tsourakakis}, {and}
  \bibinfo{person}{S.~C. Xu}.} \bibinfo{year}{2015}\natexlab{}.
\newblock \showarticletitle{Scalable Large Near-Clique Detection in Large-Scale
  Networks via Sampling}. In \bibinfo{booktitle}{{\em Proceedings of the 21th
  ACM SIGKDD International Conference on Knowledge Discovery and Data Mining}}
  {\em (\bibinfo{series}{KDD '15})}. \bibinfo{pages}{815--824}.
\newblock


\bibitem[\protect\citeauthoryear{Mukherjee and Tirthapura}{Mukherjee and
  Tirthapura}{2014}]%
        {Mukherjee14}
\bibfield{author}{\bibinfo{person}{A.~P. Mukherjee} {and} \bibinfo{person}{S.
  Tirthapura}.} \bibinfo{year}{2014}\natexlab{}.
\newblock \showarticletitle{Enumerating Maximal Bicliques from a Large Graph
  Using MapReduce}. In \bibinfo{booktitle}{{\em Proceedings of the 2014 IEEE
  International Congress on Big Data}} {\em (\bibinfo{series}{BIGDATACONGRESS
  '14})}. \bibinfo{pages}{707--716}.
\newblock
\showISBNx{978-1-4799-5057-7}


\bibitem[\protect\citeauthoryear{Newman}{Newman}{2001a}]%
        {Newman01a}
\bibfield{author}{\bibinfo{person}{M.~E.~J. Newman}.}
  \bibinfo{year}{2001}\natexlab{a}.
\newblock \showarticletitle{Scientific collaboration networks. I. Network
  construction and fundamental results}.
\newblock \bibinfo{journal}{{\em Phys. Rev. E\/}}  \bibinfo{volume}{64}
  (\bibinfo{year}{2001}), \bibinfo{pages}{016131}.
\newblock
Issue 1.


\bibitem[\protect\citeauthoryear{Newman}{Newman}{2001b}]%
        {Newman01b}
\bibfield{author}{\bibinfo{person}{M.~E.~J. Newman}.}
  \bibinfo{year}{2001}\natexlab{b}.
\newblock \showarticletitle{Scientific collaboration networks. II. Shortest
  paths, weighted networks, and centrality}.
\newblock \bibinfo{journal}{{\em Phys. Rev. E\/}}  \bibinfo{volume}{64}
  (\bibinfo{year}{2001}), \bibinfo{pages}{016132}.
\newblock
Issue 1.


\bibitem[\protect\citeauthoryear{Newman}{Newman}{2001c}]%
        {Newman2001}
\bibfield{author}{\bibinfo{person}{M.~E.~J. Newman}.}
  \bibinfo{year}{2001}\natexlab{c}.
\newblock \showarticletitle{The structure of scientific collaboration
  networks}.
\newblock \bibinfo{journal}{{\em Proceedings of the National Academy of
  Sciences\/}} \bibinfo{volume}{98}, \bibinfo{number}{2}
  (\bibinfo{year}{2001}), \bibinfo{pages}{404--409}.
\newblock


\bibitem[\protect\citeauthoryear{Opsahl}{Opsahl}{2013}]%
        {Opsahl13}
\bibfield{author}{\bibinfo{person}{T. Opsahl}.}
  \bibinfo{year}{2013}\natexlab{}.
\newblock \showarticletitle{Triadic closure in two-mode networks: Redefining
  the global and local clustering coefficients}.
\newblock \bibinfo{journal}{{\em Social Networks\/}} \bibinfo{volume}{35},
  \bibinfo{number}{2} (\bibinfo{year}{2013}), \bibinfo{pages}{159 -- 167}.
\newblock


\bibitem[\protect\citeauthoryear{Robins and Alexander}{Robins and
  Alexander}{2004}]%
        {Robins04}
\bibfield{author}{\bibinfo{person}{G. Robins} {and} \bibinfo{person}{M.
  Alexander}.} \bibinfo{year}{2004}\natexlab{}.
\newblock \showarticletitle{Small Worlds Among Interlocking Directors: Network
  Structure and Distance in Bipartite Graphs}.
\newblock \bibinfo{journal}{{\em Computational {\&} Mathematical Organization
  Theory\/}} \bibinfo{volume}{10}, \bibinfo{number}{1} (\bibinfo{year}{2004}),
  \bibinfo{pages}{69--94}.
\newblock


\bibitem[\protect\citeauthoryear{Saito and Yamada}{Saito and Yamada}{2006}]%
        {Saito06}
\bibfield{author}{\bibinfo{person}{K. Saito} {and} \bibinfo{person}{T.
  Yamada}.} \bibinfo{year}{2006}\natexlab{}.
\newblock \showarticletitle{Extracting Communities from Complex Networks by the
  k-dense Method}. In \bibinfo{booktitle}{{\em IEEE International Conf. on Data
  Mining Workshops, ICDMW}}. \bibinfo{pages}{300--304}.
\newblock


\bibitem[\protect\citeauthoryear{Sar{\i}y\"{u}ce and Pinar}{Sar{\i}y\"{u}ce and
  Pinar}{2016}]%
        {Sariyuce17}
\bibfield{author}{\bibinfo{person}{A.~E. Sar{\i}y\"{u}ce} {and}
  \bibinfo{person}{A. Pinar}.} \bibinfo{year}{2016}\natexlab{}.
\newblock \showarticletitle{Fast Hierarchy Construction for Dense Subgraphs}.
\newblock \bibinfo{journal}{{\em Proc. VLDB Endow.\/}} \bibinfo{volume}{10},
  \bibinfo{number}{3} (\bibinfo{date}{Nov.} \bibinfo{year}{2016}),
  \bibinfo{pages}{97--108}.
\newblock
\showISSN{2150-8097}


\bibitem[\protect\citeauthoryear{Sar{\i}y\"{u}ce and Pinar}{Sar{\i}y\"{u}ce and
  Pinar}{2017}]%
        {longer}
\bibfield{author}{\bibinfo{person}{A.~E. Sar{\i}y\"{u}ce} {and}
  \bibinfo{person}{A. Pinar}.} \bibinfo{year}{2017}\natexlab{}.
\newblock \showarticletitle{Peeling Bipartite Networks for Dense Subgraph
  Discovery}.
\newblock \bibinfo{journal}{{\em CoRR\/}}  \bibinfo{volume}{1611.02756}
  (\bibinfo{year}{2017}).
\newblock
\newblock
\shownote{(Extended version).}


\bibitem[\protect\citeauthoryear{Sar{\i}y{\"{u}}ce, Seshadhri, P{\i}nar, and
  {\c{C}}ataly{\"{u}}rek}{Sar{\i}y{\"{u}}ce et~al\mbox{.}}{2015}]%
        {Sariyuce15}
\bibfield{author}{\bibinfo{person}{A.~E. Sar{\i}y{\"{u}}ce},
  \bibinfo{person}{C. Seshadhri}, \bibinfo{person}{A. P{\i}nar}, {and}
  \bibinfo{person}{{\"{U}}.~V. {\c{C}}ataly{\"{u}}rek}.}
  \bibinfo{year}{2015}\natexlab{}.
\newblock \showarticletitle{Finding the Hierarchy of Dense Subgraphs Using
  Nucleus Decompositions}. In \bibinfo{booktitle}{{\em Proc. of the
  International Conf. on World Wide Web (WWW)}}. \bibinfo{pages}{927--937}.
\newblock


\bibitem[\protect\citeauthoryear{Seidman}{Seidman}{1983}]%
        {Seidman83}
\bibfield{author}{\bibinfo{person}{S.~B. Seidman}.}
  \bibinfo{year}{1983}\natexlab{}.
\newblock \showarticletitle{Network structure and minimum degree}.
\newblock \bibinfo{journal}{{\em Social Networks\/}} \bibinfo{volume}{5},
  \bibinfo{number}{3} (\bibinfo{year}{1983}), \bibinfo{pages}{269--287}.
\newblock


\bibitem[\protect\citeauthoryear{Sim, Li, Gopalkrishnan, and Liu}{Sim
  et~al\mbox{.}}{2009}]%
        {Sim09}
\bibfield{author}{\bibinfo{person}{K. Sim}, \bibinfo{person}{J. Li},
  \bibinfo{person}{V. Gopalkrishnan}, {and} \bibinfo{person}{G. Liu}.}
  \bibinfo{year}{2009}\natexlab{}.
\newblock \showarticletitle{Mining maximal quasi-bicliques: Novel algorithm and
  applications in the stock market and protein networks}.
\newblock \bibinfo{journal}{{\em Statistical Analysis and Data Mining\/}}
  \bibinfo{volume}{2}, \bibinfo{number}{4} (\bibinfo{year}{2009}),
  \bibinfo{pages}{255--273}.
\newblock


\bibitem[\protect\citeauthoryear{Tsourakakis}{Tsourakakis}{2015}]%
        {Tsourakakis15}
\bibfield{author}{\bibinfo{person}{C. Tsourakakis}.}
  \bibinfo{year}{2015}\natexlab{}.
\newblock \showarticletitle{The K-clique Densest Subgraph Problem}. In
  \bibinfo{booktitle}{{\em Proc. of the 24th International Conf. on World Wide
  Web}} {\em (\bibinfo{series}{WWW '15})}. \bibinfo{pages}{1122--1132}.
\newblock
\showISBNx{978-1-4503-3469-3}


\bibitem[\protect\citeauthoryear{Tsourakakis, Pachocki, and
  Mitzenmacher}{Tsourakakis et~al\mbox{.}}{2017}]%
        {Babis17}
\bibfield{author}{\bibinfo{person}{C.~E. Tsourakakis}, \bibinfo{person}{J.
  Pachocki}, {and} \bibinfo{person}{M. Mitzenmacher}.}
  \bibinfo{year}{2017}\natexlab{}.
\newblock \showarticletitle{Scalable Motif-aware Graph Clustering}. In
  \bibinfo{booktitle}{{\em Proceedings of the 26th International Conference on
  World Wide Web}} {\em (\bibinfo{series}{WWW '17})}.
  \bibinfo{pages}{1451--1460}.
\newblock
\showISBNx{978-1-4503-4913-0}


\bibitem[\protect\citeauthoryear{Verma and Butenko}{Verma and Butenko}{2012}]%
        {Verma12}
\bibfield{author}{\bibinfo{person}{A. Verma} {and} \bibinfo{person}{S.
  Butenko}.} \bibinfo{year}{2012}\natexlab{}.
\newblock \showarticletitle{Network clustering via clique relaxations: {A}
  community based approach}. In \bibinfo{booktitle}{{\em Graph Partitioning and
  Clustering, {DIMACS} Workshop}}. \bibinfo{pages}{129--140}.
\newblock


\bibitem[\protect\citeauthoryear{Wolf, Klinvex, and Dunlavy}{Wolf
  et~al\mbox{.}}{2016}]%
        {Wolf16}
\bibfield{author}{\bibinfo{person}{M.~M. Wolf}, \bibinfo{person}{A.~M.
  Klinvex}, {and} \bibinfo{person}{D.~M. Dunlavy}.}
  \bibinfo{year}{2016}\natexlab{}.
\newblock \showarticletitle{Advantages to Modeling Relational Data using
  Hypergraphs versus Graphs}. In \bibinfo{booktitle}{{\em {IEEE} High
  Performance Extreme Computing Conference, {HPEC}}}.
\newblock


\bibitem[\protect\citeauthoryear{Zhang and Parthasarathy}{Zhang and
  Parthasarathy}{2012}]%
        {Zhang12}
\bibfield{author}{\bibinfo{person}{Y. Zhang} {and} \bibinfo{person}{S.
  Parthasarathy}.} \bibinfo{year}{2012}\natexlab{}.
\newblock \showarticletitle{Extracting Analyzing and Visualizing Triangle
  K-Core Motifs Within Networks}. In \bibinfo{booktitle}{{\em Proc. of the IEEE
  International Conf. on Data Engineering}} {\em (\bibinfo{series}{ICDE})}.
  \bibinfo{pages}{1049--1060}.
\newblock


\end{thebibliography}

\end{document}